\documentclass[prb,twocolumn,showpacs,amsmath,amssymb,superscriptaddress]{revtex4}

\usepackage{graphicx}
\usepackage{ifthen}
\usepackage{multirow}
\usepackage{dcolumn}
\usepackage{threeparttable}
\usepackage{color} 
\usepackage{hyperref}
\hypersetup{backref=true,
 pdfnewwindow=true, colorlinks=true,
 linkcolor=blue, anchorcolor=blue,
 citecolor=blue, filecolor=blue,
 menucolor=blue, urlcolor=blue}

\def\pitem[#1]{\textsuperscript{#1\hspace{0.05cm}}\ignorespaces}
\def\pnote[#1]{\textsuperscript{#1}}

\newcommand\mcc[1]{\multicolumn{1}{c}{#1}}

\begin{document}

\title{Si-compatible candidates for \hk\ dielectrics with the Pbnm
  perovskite structure}

\author{Sinisa Coh} 
\email{sinisa@physics.rutgers.edu} 
\affiliation{
  Department of Physics and Astronomy, Rutgers University, Piscataway,
  NJ 08854-8019, USA
}
\author{Tassilo Heeg}
\affiliation{
  Department of Materials Science and Engineering, 
  Cornell University, 
  Ithaca, NY 14853, USA
}
\author{J. H. Haeni} 
\affiliation{
  Department of Materials Science and Engineering, 
  Pennsylvania State University,
  University Park, PA 16802, USA
}
\author{M. D. Biegalski}
\affiliation{
  Center for Nanophase Materials Science,
  Oak Ridge National Laboratory,
  Oak Ridge, TN 37830, USA
}
\author{J. Lettieri}
\altaffiliation[]{Deceased.}
\affiliation{
  Department of Materials Science and Engineering, 
  Pennsylvania State University,
  University Park, PA 16802, USA
}
\author{L. F. Edge}
\affiliation{
  Department of Materials Science and Engineering, 
  Pennsylvania State University,
  University Park, PA 16802, USA
}
\author{K. E. O'Brien}
\affiliation{
  Department of Materials Science and Engineering, 
  Pennsylvania State University,
  University Park, PA 16802, USA
}
\author{M. Bernhagen}
\affiliation{
  Leibniz Institute for Crystal Growth, 
  Max-Born-Stra\ss e 2, D-12489 Berlin
  (Adlershof), Germany
}
\author{P. Reiche}
\affiliation{
  Leibniz Institute for Crystal Growth, 
  Max-Born-Stra\ss e 2, D-12489 Berlin
  (Adlershof), Germany
}
\author{R. Uecker}
\affiliation{
  Leibniz Institute for Crystal Growth, 
  Max-Born-Stra\ss e 2, D-12489 Berlin
  (Adlershof), Germany
}
\author{S. Trolier-McKinstry}
\affiliation{
  Department of Materials Science and Engineering, 
  Pennsylvania State University,
  University Park, PA 16802, USA
}
\author{Darrell G. Schlom}
\affiliation{
  Department of Materials Science and Engineering, 
  Cornell University, 
  Ithaca, NY 14853, USA
}
\author{David Vanderbilt} 
\affiliation{
  Department of Physics and Astronomy, Rutgers University, Piscataway,
  NJ 08854-8019, USA
}

\date{\today}

\def\cub{$Pm\bar{3}m$} 
\def\cubp{$O_h^1$}     
\def\pbnm{$Pbnm$}      
\def\pbnmB{$\bf Pbnm$} 
\def\pbnmp{$D_{2h}^{16}$}
\def\rfe3c{$R3c$}        
\def\Irbar3c{$\it R\bar{3}c$}
\def\rbar3c{$R\bar{3}c$}     
\def\rs{$P2_1/c$}      
\def\rsp{$C_{2h}^{5}$} 
\def\3{$_3$}
\def\bpr{{\it B}$'$}
\def\labbo{La$_2${\it B}\bpr{}O$_6$}
\def\xe{\epsilon_{xx}}
\def\ye{\epsilon_{yy}}
\def\ze{\epsilon_{zz}}
\def\xze{\epsilon_{xz}}
\def\1e{\Delta\epsilon_{\parallel}}
\def\2e{\Delta\epsilon_{\perp}}
\def\ae{\bar{\epsilon}}
\def\thM{\theta_{\rm M}}
\def\thR{\theta_{\rm R}}
\def\G{\Gamma}
\def\dsum{\displaystyle\sum}
\def\aac{($a^-a^-c^+$)}
\def\aaap{($a^-a^-a^+$)}
\def\aaam{($a^-a^-a^-$)}
\def\P{{\bf P}}
\def\eion{\epsilon^{\rm ion}}
\def\hk{high-K} 
\def\hki{high-{\it K}}

\pacs{77.22.-d,77.55.df,85.50.-n}

\begin{abstract}  
  We analyze both experimentally (where possible) and theoretically
  from first-principles the dielectric tensor components and crystal
  structure of five classes of \pbnm\ perovskites. All of these
  materials are believed to be stable on silicon and are therefore
  promising candidates for \hki\ dielectrics. We also analyze the
  structure of these materials with various simple models,
  decompose the lattice contribution to the dielectric tensor into
  force constant matrix eigenmode contributions, explore
  a peculiar correlation between structural and dielectric
  anisotropies in these compounds and give phonon frequencies and
  infrared activities of those modes that are infrared-active. 
  We find that CaZrO\3, SrZrO\3,
  LaHoO\3, and LaYO\3 are among the most promising candidates
  for \hki\ dielectrics among the compounds we considered.
\end{abstract}

\maketitle

\section{Introduction}
\label{sec:intro}

As a result of the ongoing down-scaling of complementary
metal-oxide-semiconductor (CMOS) integrated circuits, the SiO$_2$ gate
oxide of field effect transistors is getting thinner and thinner in
every new generation of devices.  \cite{moore} Therefore the leakage
current due to quantum-mechanical tunneling through the dielectric
interface is increasing. One way to reduce this current is to replace
SiO$_2$ with a material that has a higher dielectric constant.
Such a \hki\ dielectric layer with the same effective dielectric
thickness (i.e., providing the same capacitance) could be physically
thicker and thus reduce the gate leakage.

In order for this replacement material to be useful in practical
applications on silicon, it also needs to be stable in contact with
silicon up to $\sim$1000~$^{\circ}$C, and among other things it must
also have an appropriate band alignment with silicon.
\cite{ogale-oxide,schlom-mrs,robertson}
Currently, a hafnia-based dielectric 
is used as a replacement to SiO$_2$ in advanced CMOS transistors in
production.~\cite{mistry,intel-highk-2,scansen}
There are, however, drawbacks to this material too, e.g., the 
limited {\it K} that it provides and undesirable threshold
voltage shifts arising from highly mobile
oxygen vacancies.~\cite{guha}
This brings up the natural question: {\it
which other materials exist that would satisfy these requirements
and would enable the scaling of MOSFETs to continue beyond today's
hafnia-based dielectrics}?

The stability of single component oxides on silicon has been
demonstrated both experimentally and from thermodynamic
analysis,\cite{schlom-mrs} and a candidate list of
multicomponent oxide materials has been compiled.~\cite{ogale-oxide} A
promising group of these materials consists of perovskite oxides having a
\pbnm\ (or closely related \rs) space group.
\begin{figure*}[!tb]
  \includegraphics[width=17.9cm]{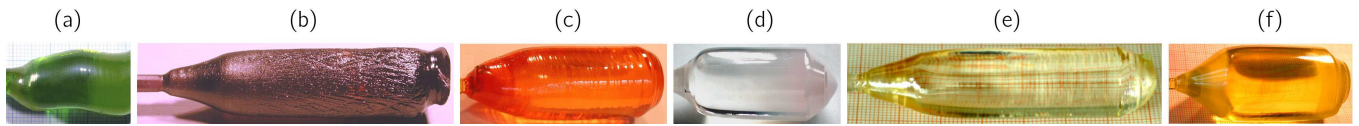}
  \caption{\label{fig:cry} Photographs of single crystal rare-earth
    scandates
    grown along the $[110]$ direction by the Czochralski method.
    (a) PrScO\3\ with a diameter of 12~mm,
    (b) NdScO\3\ with a diameter of 16~mm, (c) SmScO\3\ with 
    a diameter of
    18~mm, (d) GdScO\3\ with a diameter of 32~mm, (e) TbScO\3\
    with a diameter of 18~mm, and (f) DyScO\3\ with
    a diameter of 32~mm.}
\end{figure*}
These compounds are at the focus of the present work. Some of them have been
studied in thin-film form,
\cite{christen,afanasev-1,zhao,heeg,edge-1,cicerrella,afanasev-2,edge-2,
  sivasubramani,lopes,wang,edge-3,ozben,roeckerath,wagner,myllymaki,
  kim,thomas,heeg-2} but the full dielectric tensor of these materials
has not yet been established, making the selection of materials best
suited for \hki\ applications difficult.
Some of these materials could also be of interest for microwave
dielectric applications.\cite{wersing,terrell}
Thus we decided to study,
both theoretically and experimentally, the structural and dielectric
properties of these compounds. The calculations are carried out using
density-functional theory, and we compare the results with experimental
data where we could obtain suitable samples for measurements.  To our
knowledge, previous theoretical calculations have been carried out
in only a few cases.\cite{delugas-th, vali-zr, vali-hf}

The paper is organized as follows.
Explanations of both the experimental and theoretical methods used in this
work are given in Sec.~\ref{sec:pre}.  The main results on the
structural and dielectric properties are given and discussed in 
Sec.~\ref{sec:res}. 
There we also discuss the correlations between the structural and dielectric
properties of these perovskites, decompose the ionic contribution of the
dielectric tensor into components arising from various force constant matrix
eigenvectors, and discuss the effect of 
{\it B}$_{\it A}$ antisite defects on
the dielectric properties.
We finish with a brief summary in Sec.~\ref{sec:sum}.

Supplementary material~\cite{EPAPS_arxiv} contains the results of our
calculations of the zone-center phonon frequencies, as well as the
infrared activities for those modes that are infrared-active.

\subsection{Compounds under consideration}
\label{sec:comp}

In this work, we consider the following five groups of 
perovskites having the \pbnm\ space group.

The first group are rare-earth scandates having formula {\it A}ScO\3\
where {\it A} is a rare-earth atom. In
Sec.~\ref{sec:resc_die} we report experimental measurements of
the full dielectric tensors for PrScO\3, NdScO\3, SmScO\3, GdScO\3\ and
DyScO\3. Calculations were done on these and also on LaScO\3\ and TbScO\3.
Note that
HoScO\3, ErScO\3, TmScO\3, YbScO\3, LuScO\3, and YScO\3 do not form
single crystals with the perovskite structure from the melt at
atmospheric pressure.  Rather, they form solid solutions of {\it
  A}$_2$O\3\ and Sc$_2$O\3, i.e., ({\it A},Sc)$_2$O$_3$, with the
bixbyite structure.\cite{badie-1,badie-2,coutures} Nevertheless,
LuScO\3~\cite{heeg-2} and YbScO\3~\cite{schubert} have been formed in
perovskite form as thin films via epitaxial stabilization, and others
might be made in the same way.  To analyze trends within this group of
compounds, we also did the calculations of dielectric tensors on
LuScO\3\ and YScO\3\
in the \pbnm\ perovskite structure;
see Sec.~\ref{sec:resc_die} for the details.

The second group consists of rare-earth yttrates with formula {\it A}YO\3.
Only one such compound,
LaYO\3, is known to form a perovskite,\cite{berndt} but to analyze
trends the dielectric tensor of DyYO\3\ in the perovskite structure was
also calculated.

In the third group we consider CaZrO\3, SrZrO\3\ and SrHfO\3\
perovskites.  Experimentally, we find that SrZrO\3\ and SrHfO\3\ do
not form single crystals, but instead are rather heavily twinned.

The fourth group of compounds have the formula \labbo\ where the
{\it B} atom is either Mg or Ca and \bpr\ is either Zr or Hf.
Little is known experimentally
about these compounds, and single crystals of these
compounds have not been made.\cite{labbo}

The last group of compounds we considered have the formula {\it AA}'O\3,
where both {\it A} and {\it A}' are rare-earth atoms.  These include
the 11 of such compounds that are known to form the perovskite structure
with space group \pbnm\ at atmospheric pressure: LaHoO\3, LaErO\3,
LaTmO\3, LaYbO\3, LaLuO\3, CeTmO\3, CeYbO\3, CeLuO\3, PrYbO\3,
PrLuO\3, and NdLuO\3.\cite{berndt,ito,bharathy} We calculated the
dielectric and structural properties of all of these compounds.
The experimental determination, however, of the dielectric tensor in
this group of compounds was done only for LaLuO\3; the results
will be published elsewhere.~\cite{heeg-apl}

\section{Preliminaries}
\label{sec:pre}

\subsection{Structure of \pbnmB\ perovskites}
\label{sec:pbnm}

\begin{figure}[!h]
  \includegraphics{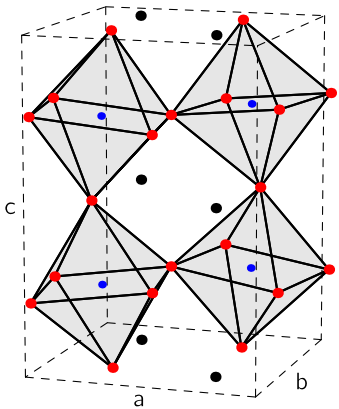}
  \caption{\label{fig:str} (Color online)
    20-atom primitive cell of a {\pbnm}-distorted
    {\it AB}O\3\ perovskite. {\it A}-site atoms are shown in
    black, {\it B}-site atoms in blue (at the centers of the
    octahedra) and oxygen atoms in red (at the vertices of the
    octahedra). Orthorhombic unit cell vectors ($a$,
    $b$, and $c$) are also indicated.}
\end{figure}

\begin{figure}[!h]
  \includegraphics{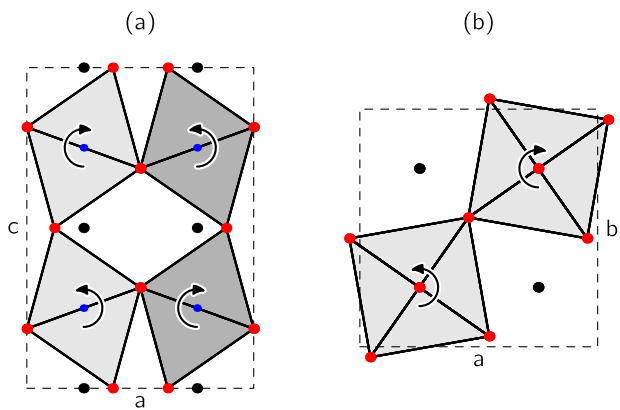}
  \caption{\label{fig:rot} (Color online)  (a) Projection on the
    $a$-$c$ plane of the structure with $\thR>0$ and $\thM=0$.
    (b) Projection on the $a$-$b$ plane for $\thR=0$ and
    $\thM>0$. Color coding of atoms and positions of axes labels are
    the same as in Fig.~\ref{fig:str}.  In (a), a darker shading is
    used to indicate the two octahedra that are further away along the
    $y$-coordinate; in (b), the two bottom octahedra are exactly
    below the two top octahedra.}
\end{figure}

The ideal cubic perovskite {\it AB}O\3\ consists of a network of
corner-shared octahedra, each with an oxygen on its vertices and a {\it
  B} atom at its center, and {\it A} ions that are 12-fold coordinated
in the spaces between octahedra.  It is well
known that perovskites having sufficiently small {\it A}-site
ions (i.e., a small Goldschmidt tolerance
factor\cite{goldschmidt,woodward-oct}) often allow for a distorted
perovskite structure that has a rotated framework of oxygen octahedra
and displaced {\it A}-site ions. This lowers the space group symmetry
from cubic \cub\ (\cubp) to orthorhombic \pbnm\ (\pbnmp), and the
number of {\it AB}O\3\ formula units per primitive cell increases from
$Z=1$ to $Z=4$, as shown in Fig.~\ref{fig:str}.
The rotations of the octahedra in the \pbnm\ space group
can be decomposed into two steps. The first step is the rotation
around the $[110]$ direction of the original cubic frame (the cubic
frame is rotated by 45$^{\circ}$ around the $z$ axis with respect to
the \pbnm\ frame) by an angle $\thR$ as in Fig.~\ref{fig:rot}(a), and
the second step is a rotation around $[001]$ by $\thM$ as in
Fig.~\ref{fig:rot}(b). The rotations must be done in that order to
prevent distortions of the octahedra.  The
pattern of neighboring octahedral rotations is denoted by \aac\ in
Glazer notation \cite{glazer-oct} (or see the directions of the arrows in
Fig.~\ref{fig:rot}). These rotations also allow for the displacement
of {\it A}-site ions in the $x$-$y$ plane without further lowering of the
space-group symmetry.

\subsection{Computational methods}
\label{sec:cmpMet}

The main computational method we are using is the density-functional
theory as implemented in the Quantum-Espresso package.\cite{QE-2009}
The exchange-correlation functional was approximated using a
generalized gradient approximation (GGA) of the Perdew-Burke-Ernzerhof
type \cite{perdew-pbe} and ultrasoft pseudopotentials were
employed.\cite{vanderbilt-uspp} The electronic wavefunctions were
expanded in a basis of plane waves with kinetic energy up to 40~Ry,
while the charge density was expanded up to 300~Ry.  The Brillouin
zone was sampled using a $4\times 4\times 4$ Monkhorst-Pack
grid.\cite{monkhorst-grid}

A new set of ultrasoft pseudopotentials\cite{vanderbilt-uspp}
for the lanthanoide series of rare earths, from La to Lu, were generated
for the present project.  In all cases the $f$-shell filling was
chosen as appropriate for the 3+ valence state: one $f$ electron for
Ce, two for Pr, etc.  The $f$ electrons were then considered to be
in the core (and un-spin-polarized) for the proposes of generating
the pseudopotentials.  Thus, the $f$ electrons are not explicitly
included in the solid-state calculations.  Such an approximation can
be justified whenever the strong on-site Coulomb interactions of
electrons in the $f$ shell drive the occupied $f$
states well below, and the unoccupied states well above, the energy
range of interest for $spd$ bonding in the crystal.  Of course, this
will not be a good approximation for some heavy-fermion or
mixed-valent systems, and in any case our approach is obviously unable
to describe phenomena involving magnetic ordering of $f$ electrons at
low temperature.  Nevertheless, we believe that this approach is quite
reasonable for the present purposes.

The artificial nature of the scattering in the $f$ channel did, however,
pose some problems in the pseudopotential construction.  In particular, we
found that the lattice constant of a perovskite containing the rare-earth
atom in question could differ for two pseudopotentials having different
scattering properties in the $f$ channel; this causes problems since
the usual approach of matching to the all-electron $f$ scattering is
not appropriate in the present case.  To ameliorate this problem, the
$f$-channel parameters of these pseudopotentials were optimized so that
resulting pseudopotentials would give the ``correct'' cell volumes for
simple rare-earth compounds.  Since the GGA typically overestimates
crystal volumes by about 1-2\%,\cite{gga-volume} the optimization
was actually done in order to produce a corresponding overestimate
in a consistent fashion.

\begin{figure}[!h]
  \includegraphics{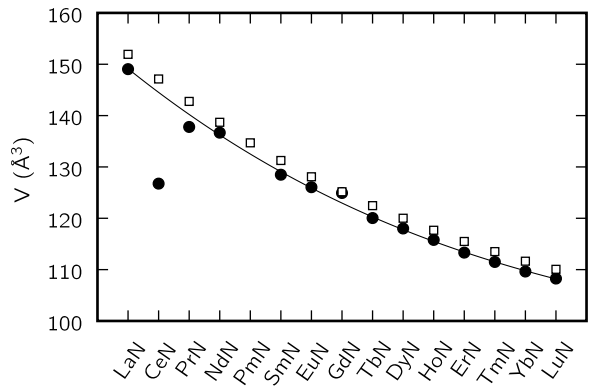}
  \caption{\label{fig:psp} Unit cell volumes of all rock-salt
    rare-earth nitrides, in \AA$^3$. Empty squares are the results
    obtained using our optimized ultrasoft pseudopotentials for
    rare-earth atoms; solid circles are
    experimental\cite{landolt-nitrides} results. The solid line is
    a fit to the experimental values excluding CeN, PrN, and GdN.}
\end{figure}

For this procedure, our compounds of choice
were the rare-earth nitrides with the simple rock-salt structure. The
experimental\cite{landolt-nitrides} and calculated volumes of these
nitrides are indicated in Fig.~\ref{fig:psp}.
Note that the volumes of CeN, PrN, and GdN show an anomalous behavior
that is presumably due to strong correlation effects associated with
the proximity to a mixed-valent regime,~\cite{gdn} and
therefore they will not be correctly treated by our GGA calculation.
To avoid this problem we first carried out a smoothened fit of the
experimental volumes versus atomic number over the lanthanoide
nitride series, but with CeN, PrN, and GdN omitted from the fit, as
shown by the solid line in Fig.~\ref{fig:psp}.  We then
used these fitted values to set the target volumes for the optimization
of the pseudopotentials.

We used density-functional perturbation theory\cite{baroni-rmp} to
calculate the dielectric response. Both purely electronic
$\epsilon^{\rm el}$ and ionic $\epsilon^{\rm ion}$ contributions were
calculated.\cite{cockayne} The electronic part is defined as
\begin{align}
\epsilon_{\alpha\beta}^{\rm el} = \delta_{\alpha\beta} + 4 \pi
\frac{\partial P_{\alpha}}{\partial {\cal E}_{\beta}}\bigg\vert_{u=0},
\end{align}
where $P_{\alpha}$ is the polarization induced by the electric field
${\cal E}_{\beta}$ while all ions are held fixed ($u=0$).  The
remaining component of the dielectric response is by definition
the ionic contribution $\epsilon^{\rm ion}$.

This ionic part can be calculated from the force-constant matrix
$\Phi_{i\alpha,j\beta}$ and the Born effective charge matrix
$Z_{i,\alpha\beta}$. The force-constant matrix is defined as
\begin{align}
  \Phi_{i\alpha,j\beta}=\frac{\partial^2 E}{\partial u_{i\alpha}
    \partial u_{j\beta}},
\end{align}
where $E$ is the total energy of the system and $u_{i\alpha}$ is the
displacement of the $i$-th atom along the direction $\alpha$.  We will
denote the $n$-th normalized eigenvector of this matrix as
$\xi^n_{i\alpha}$ and its eigenvalue as $\mu_n$.
The Born effective
charge matrix is defined as
\begin{align}
  Z_{i,\alpha\beta}=\frac{V}{e} \frac{\partial P_{\alpha}}{\partial
    u_{i\beta}},
\end{align}
where $P_{\alpha}$ is the polarization induced in a crystal by the
displacement of the $i$-th atom in the direction $\beta$. $V$ is the
volume of the unit cell and $e$ is the electron charge.
Finally, the ionic part of the dielectric tensor can be written as
\begin{align}
  \eion_{\alpha\beta}=\frac{4 \pi e^2}{V} \dsum_{n}
  \frac{1}{\mu_n} Q_{\alpha}^n Q_{\beta}^n,
  \label{eq:eion}
\end{align}
where the charge $Q_{\alpha}^n$ of the $n$-th eigenmode is defined
through the effective charge matrix as $Q_{\alpha}^n=\dsum_{i\beta}
Z_{i,\alpha\beta} \xi_{i\beta}^n$.

\subsection{Experimental methods}
\label{sec:expMet}

\subsubsection{Crystal growth}

PrScO\3, NdScO\3, SmScO\3, GdScO\3, TbScO\3, and DyScO\3\
single crystals were grown using an automated Czochralski technique
with RF-induction heating.~\cite{uecker-1,uecker-2} Pre-dried powders
of Sc$_2$O$_3$, Pr$_6$O$_{11}$, Nd$_2$O$_3$, Sm$_2$O$_3$,
Gd$_2$O$_3$, Tb$_4$O$_7$, and Dy$_2$O$_3$, were mixed in the
stoichiometric ratio, pressed, and sintered at about 1400~$^{\circ}$C
for 15 h.  Due to the high melting temperature of PrScO$_3$,
NdScO$_3$, SmScO$_3$, GdScO$_3$, TbScO$_3$, and DyScO$_3$
($\sim$2100~$^{\circ}$C), a crucible (cylindrical with 40~mm 
or 60~mm diameter
and 40~mm or 60~mm height, depending on the crystal diameter) 
and an active afterheater made of iridium were used.
Flowing nitrogen or argon was used as the growth atmosphere.  
Those of these rare-earth scandates for which the radiative heat
transport via the crystal is hindered by absorption suffer from a
serious problem of bulk crystal growth, namely that they tend to
exhibit a spiral growth which distinctly decreases the yield.~\cite{uecker-1}
Due to
the lack of seed crystals, the initial growth experiments were
performed with an iridium seed rod. Because these materials tend to
grow as large single-crystalline grains, suitable seeds could be
selected at a very early stage. 

All rare-earth scandate crystals were grown along the 
$\left[110\right]$ orientation.
The pulling rate was $0.8$-$2$~mm$\cdot$h$^{-1}$ 
and the rotation rate was $8$-$15$~min$^{-1}$
(depending on the crystal diameter). The crystals were 35-50~mm in
length and 18 or 32~mm in diameter. The PrScO\3\ crystals have a green
color, the NdScO\3\ crystals are dark purple,
the SmScO\3\ and DyScO\3\ crystals are light yellow,
the GdScO\3\ crystals are colorless, and the TbScO\3\ crystals
are nearly colorless (see Fig.~\ref{fig:cry}).

In addition
to the Czochralski growth of PrScO$_3$, NdScO$_3$, SmScO$_3$,
GdScO$_3$, TbScO$_3$, and DyScO$_3$, the single-crystal
growth of HoScO$_3$, YScO$_3$, and solid solutions of rare-earth
scandates that would be expected to have smaller lattice constants
than DyScO\3 (e.g., Dy$_{0.5}$Lu$_{0.5}$ScO\3),
was also attempted using the floating-zone technique. In all cases,
pre-dried powders of the rare-earth oxides ({\it R}$_2$O$_3$)
and Sc$_2$O$_3$ were ground,
pressed into a rod, and sintered at 1400~$^{\circ}$C for 6 hours. The
floating-zone machine was a Gero (model SPO), and heating was
performed with two bulbs at the focal point of ellipsoidal mirrors.
Unfortunately, the 1~kW
rated output power of the quartz-halogen lamps used in this
floating-zone system was close to the melting temperature of the compounds,
limiting the size of the crystals that could be grown to $<2$~mm
diameter.
SrZrO\3, SrHfO\3, and LaScO\3\ single crystals grown by the floating
zone technique were also studied.~\cite{srzro}

The available {\it R}$_2$O$_3$-Sc$_2$O$_3$ binary phase diagrams show a
distinct transition between the Dy$_2$O$_3$-Sc$_2$O$_3$ and the
Ho$_2$O$_3$-Sc$_2$O$_3$ systems (Dy and Ho are neighboring elements
in the periodic table).
According to published phase diagrams, the rare-earth elements of La,
Nd, Sm, Gd, and Dy all form {\it R}ScO$_3$
compounds that melt congruently. In contrast
the oxides of Ho, Er, Tm, Yb, and Lu form complete
solid solutions with Sc$_2$O$_3$ at their melting
points.~\cite{badie-1,badie-2,coutures} (The same happens with Y as
well.~\cite{badie-1,badie-2,coutures})
Although HoScO$_3$ has been synthesized with solid-state
techniques at temperatures (well below the melting point) at which the
perovskite polymorph of HoScO\3\ is stable, our results confirm
that the perovskite polymorph of HoScO\3\ is not stable at its melting
point in agreement with
existing phase diagrams. Analysis of our Y-Sc-O and Ho-Sc-O single
crystals revealed that they were solid solution mixtures of
Ho$_2$O$_3$ and Sc$_2$O$_3$, and Y$_2$O$_3$ and Sc$_2$O$_3$, not the
desired perovskite polymorphs of HoScO$_3$ and YScO$_3$. The only
compound that did not behave according to its phase diagram was
LaScO$_3$.  Attempts to grow LaScO$_3$ resulted in crystals that were
a mixture of three different phases including LaScO$_3$, Sc$_2$O$_3$,
and a third, unidentified phase.~\cite{srzro} X-ray analysis of the
post-annealed polycrystalline feed rod showed single-phase LaScO$_3$,
consistent with the existing phase diagram. The reason for the
introduction of second phases during melting is not clear and requires
more study.

\subsubsection{Electrical characterization}
\label{sec:expElec}

Because PrScO$_3$, NdScO$_3$, SmScO$_3$, GdScO$_3$, and DyScO$_3$ are
orthorhombic at room temperature, their dielectric tensor contains
three independent coefficients that can be measured along the three
principle crystal axes.~\cite{newnham} From the grown PrScO$_3$,
NdScO$_3$, SmScO$_3$, GdScO$_3$, and DyScO$_3$ single crystals, slices
were cut in different orientations for electrical characterization.
For GdScO$_3$, samples were cut from the as-grown single
crystals along the $\left[100\right]$, $\left[010\right]$, and
$\left[001\right]$ directions, and the dielectric tensor coefficients
were measured using a parallel-plate capacitor configuration. For
PrScO$_3$, NdScO$_3$, SmScO$_3$, and DyScO$_3$, several additional
orientations were prepared, so that a least-squares fit technique
could be applied using standard methods~\cite{newnham} to calculate
the dielectric tensor, using their known lattice
parameters.\cite{uecker-2,gesing}

Unfortunately, the SrZrO$_3$ and SrHfO$_3$ crystals were heavily
twinned. The formation of twins in the three pseudocubic orientations
is facilitated in these two orthorhombic systems because the deviation from
the cubic symmetry is very small.  As a result, only an average dielectric
constant could be measured.  

To obtain capacitors, gold or platinum electrodes were
evaporated onto both sides of the approximately 0.5 mm thick slabs
with areas ranging from 15 to 100 mm$^2$, and capacitance measurements
were made at room temperature
with an HP4284A using a 16034E test fixture at 10~kHz, 100~kHz, and
1~MHz. No edge capacitance corrections were used because of the large
ratio of the electrode area to the thickness of the samples. The room
temperature  dielectric loss of the samples was generally very small 
($<0.1\%$ for most samples) and the frequency dispersion was negligible in 
the measured range.
The temperature dependent dielectric measurements on GdScO\3\ and  DyScO\3\ 
were made using a HP 4284a LCR meter with a dipstick cryostat.

\section{Results and discussion}
\label{sec:res}

\subsection{Structural properties}
\label{sec:str}

We focus first on the structural properties of these systems.
The structure of the \pbnm\ perovkites is described by three orthorhombic
lattice constants plus two {\it A}-site and five oxygen Wyckoff
parameters.  Figure~\ref{fig:str_data} shows graphically
the most important structural parameters of these systems, while
Table~\ref{tab:str} gives detailed information on all the structural
parameters.  Rotational angles in Table~\ref{tab:str} and in
Fig.~\ref{fig:str_data} were calculated by fitting the structural
parameters to a model in which the octahedra are perfectly rigid (see
Sec.~\ref{sec:rigid} for the details of this model).

Overall we find good agreement with experimental values for the
structural parameters. The Wyckoff coordinates in particular are in
excellent agreement with experiments, with the average error being on the
order of $2\cdot10^{-3}$.  The volume of the unit cell, on the other
hand, is consistently overestimated by 1-2\%, as is usually expected
from the GGA exchange-correlation functional, and as we would expect
from our construction of the rare-earth pseudopotentials.

All structures show an angle $\thR$ that is about $\sqrt{2}$
times larger than $\thM$. Therefore, consecutive rotations by $\thR$
and $\thM$ can be considered approximately as a single rotation around
a $\{111\}$ axis in the cubic frame. That is, the actual \aac\ pattern
of rotations is very nearly \aaap\ in the Glazer notation.\cite{glazer-oct}
See Sec.~\ref{sec:corr} for a more
detailed discussion.

\subsubsection{Rare-earth scandates}

The rare-earth scandates {\it A}ScO\3\ show a decrease in volume by
$\sim$9\% while going along the series from {\it A}=La to {\it A}=Dy
(the calculated primitive unit cell volume is 271.40~\AA$^3$ for
LaScO\3\ and 249.81~\AA$^3$ for DyScO\3). On the other hand, the Sc-O
distance remains nearly constant along the series (2.12~\AA\ for
LaScO\3\ and 2.11~\AA\ for DyScO\3), which means that the change in
volume is almost entirely due to the larger octahedral rotation angles
for DyScO\3\ as compared to LaScO\3.
Our calculations also show that the same trend continues
all the way to LuScO\3.

\subsubsection{Rare-earth yttrates}

The rare-earth yttrates have a very similar behavior as the
rare-earth scandates. The main quantitative structural difference
between the two comes from the fact that yttrium is a larger ion
than scandium. This leads to a larger volume for the yttrates, and
also a larger rotation angle due to a smaller tolerance factor.

\subsubsection{CaZrO\3, SrZrO\3, and SrHfO\3}

SrZrO\3\ and SrHfO\3\ have quite similar structural
properties. The main difference can be traced to the fact that
Hf is a smaller ion than Zr.  Therefore, the calculated average Hf-O
distance is 2.07~\AA, while the average Zr-O distance is 2.11~\AA.
Furthermore, their octahedral rotation angles are about 1.7 times
smaller than in the rare-earth scandates.

In CaZrO\3\ the average Zr-O distance is 2.10~\AA, which is very close
to the corresponding
distance in SrZrO\3\ and SrHfO\3. Thus, the main reason why
CaZrO\3\ has a smaller volume than SrZrO\3\ is because of the larger
rotation angles in CaZrO\3.

\subsubsection{\labbo\ compounds}

We consider \labbo\ compounds with {\it B}=Mg or Ca and
\bpr=Zr or Hf.  These compounds are expected to exhibit rock-salt
ordering of the {\it B}-site ions as a result of the difference in
charge and ionic radius between the {\it B} and
\bpr\ ions.\cite{anderson-bord}  This ordering reduces
the symmetry from the orthorhombic \pbnm\ to
the monoclinic \rs\ (\rsp) space group.

\begin{figure*}
  \includegraphics{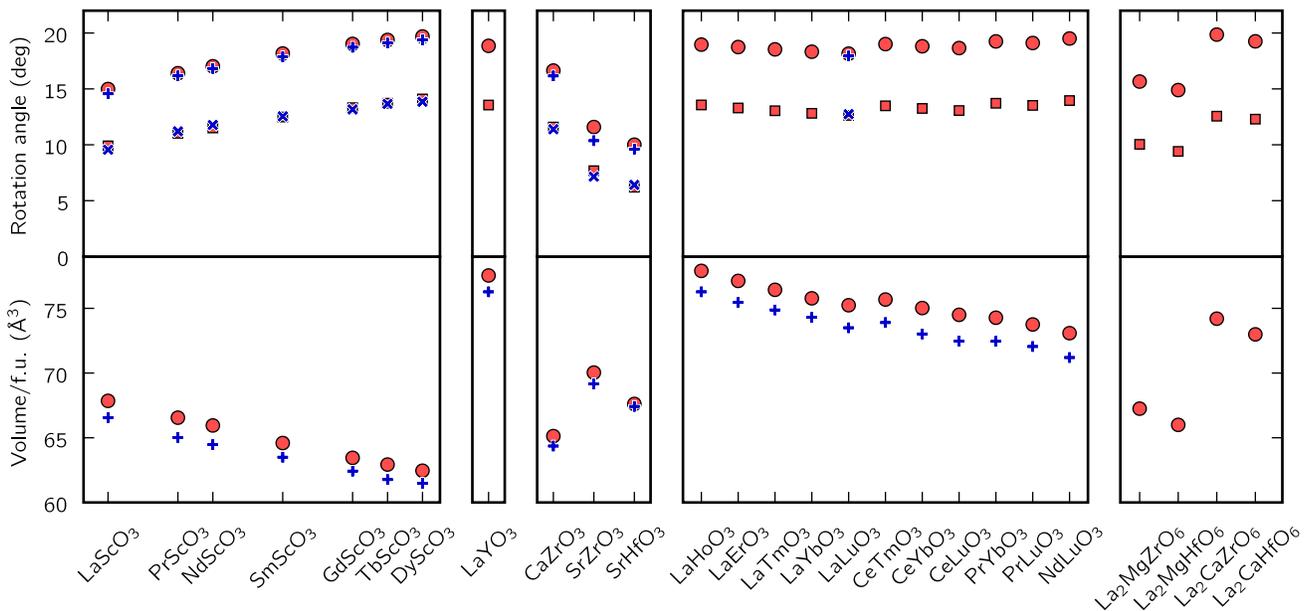}
  \caption{\label{fig:str_data} Structural information for all systems we
    considered.  Bottom pane shows volume in \AA$^3$ per formula
    unit (f. u.) of {\it AB}O\3. Calculated values are shown as red circles,
    and experimental values as blue plus symbols, if available.  Top
    pane shows oxygen octahedra rotation angles in degrees.
    Theoretical values are shown with red circles and squares and
    experimental values with blue plus and cross symbols. $\thR$
    angles are shown with circles and plus symbols while $\thM$ angles
    with squares and cross symbols. For \labbo\ systems the average of
    $\thM$ and $\thM'$ is given.  For numerical values see
    Table~\ref{tab:str} and \ref{tab:strbB}.}
\end{figure*}

The structural properties for these systems are reported in
Fig.~\ref{fig:str_data} and in Table~\ref{tab:strbB}.  The
rotational angles are obtained by a fit to the rigid-octahedra model
in which we have allowed for different sizes of {\it B}- and
\bpr-centered octahedra. (See the end of Sec.~\ref{sec:rigid}
for details.)

The unit cell volume is larger by about 5~\AA$^3$ per primitive cell
for the compounds containing Zr than for those containing Hf. On
the other hand, compounds with Ca are larger by about 28~\AA$^3$
than those containing Mg.  Similarly, the rotation angles are larger in
compounds containing Ca than in those with Mg. The discrepancy between
octahedral sizes is largest for La$_2$CaHfO$_6$ (12\% linear increase)
and smallest for La$_2$MgZrO$_6$ (0.4\% linear increase).

\subsubsection{Rare-earth rare-earth perovskites}

We now briefly analyze the structural properties of \pbnm\
perovskites of type {\it AA}'O\3\ where both {\it A} and {\it A}' are
rare-earth atoms. All eleven compounds we considered are known
experimentally to form the perovskite structure in the \pbnm\ space
group.~\cite{berndt,ito,bharathy}

Among these 11 compounds, the largest unit-cell volume of
311.58~\AA$^3$ is found in LaHoO\3, and the smallest
of 292.32~\AA$^3$ is in NdLuO\3,
Oxygen oxtahedral rotation angles are quite large in
all of these compounds and show very little variation from one
compound to another. The trends of the rotation angles are as expected
from a tolerance-factor analysis: perovskites with smaller 
{\it A}-site ions but the same {\it B}-site ions have larger oxygen
octahedral rotation angles, and the opposite is true for the
{\it B}-site ions.

\begin{table*}
  \caption{\label{tab:str} Structural parameters of the \pbnm\ perovskites
    we considered. Our calculated values are indicated with the letter T,
    experimental values with E and other theoretical data with O.
    Fitted octahedra rotation angles $\thR$ and $\thM$ are
    given in degrees, see  Sec.~\ref{sec:rigid} for the
    details. {\it A}-site ions occupy the $4c$ site and in cubic
    configuration $x_1=0$ and $y_1=1/2$. {\it B}-site ions are on the $4a$
    site. One type of oxygen sites are at $4c$ and in the cubic case $x_2=0$
    and $y_2=0$. The remaining oxygens are at $8d$ and $x_3=1/4$,
    $y_3=1/4$ and $z_3=0$ in the cubic case.}
\begin{threeparttable}
\begin{ruledtabular}
\begin{tabular*}{\textwidth}{
   ll
   D..{1.4}D..{1.4}D..{1.4}
   D..{3.2}
   D..{1.4}D..{1.4}
   D..{1.4}D..{1.4}D..{2.4}D..{2.4}D..{1.4}D..{1.4}D..{1.4}
   D..{2.1}D..{2.1}}
   & & \multicolumn{6}{c}{Unit cell parameters} & \multicolumn{7}{c}{Wyckoff
  coordinates} & \multicolumn{2}{c}{Model} \\
  \cline{3-8}\cline{9-15}\cline{16-17}
  & & \mcc{$a$} & \mcc{$b$} & \mcc{$c$} &
  \mcc{$V$} & \mcc{$b/a$} & \mcc{$c/a$} & \mcc{$x_1$} &
  \mcc{$y_1$} & \mcc{$x_2$} & \mcc{$y_2$} & \mcc{$x_3$} & \mcc{$y_3$} &
  \mcc{$z_3$} & \mcc{$\thR$} & \mcc{$\thM$}\\
  & & \mcc{(\AA)} & \mcc{(\AA)} & \mcc{(\AA)} &
  \mcc{(\AA$^3$)} & & & & & & & & & & \mcc{($^{\circ}$)} & \mcc{($^{\circ}$)}\\
 \hline
 LaScO\3\ & T & 5.7030 & 5.8414 & 8.1469 & 271.40 & 1.0243 &
 1.4285 & 0.0123 &  0.5467 & -0.0977 & -0.0323 & 0.2059 & 0.2943
 & 0.0523 & 14.9 &  9.9 \\ 
 & E\pnote[a] & 5.6803 &  5.7907 &  8.0945 &  266.25 &  1.0194
 &  1.4250 &  0.0100 &  0.5428 &  -0.0968 &  -0.0277 &  0.2073 &  0.2958
 &  0.0521 &  14.6 &  9.6 \\
 PrScO\3\ & T & 5.6372 & 5.8367 & 8.0908 & 266.21 & 1.0354 &
 1.4352 & 0.0148  & 0.5537 & -0.1049 & -0.0377 & 0.2014 &
 0.2979 & 0.0557 & 16.4 &  11.0 \\ 
 & E\pnote[j] & 5.608 &  5.780 &  8.025 &  260.1 &  1.0307 &
 1.4310 & 0.0121 & 0.5507 & -0.1052 & -0.0395 & 0.1977 & 0.3008 & 
 0.0555 &  16.2 &  11.2 \\
 NdScO\3\ & T & 5.6077 & 5.8317 & 8.0667 & 263.80 & 1.0399 &
 1.4385 & 0.0159 &  0.5562 & -0.1083 & -0.0401 & 0.1997 & 0.2992
 & 0.0574 & 17.0 &  11.5 \\ 
 & E\pnote[b] & 5.577 &  5.777 &  8.005 &  257.9 &  1.0359 &
 1.4354 &  0.0133 &  0.5532 &  -0.1088 &  -0.0418 &  0.1953 &  0.3020 &
  0.0571 &  16.8 &  11.8 \\
 SmScO\3\ & T & 5.5483 & 5.8067 & 8.0196 & 258.37 & 1.0466 &
 1.4454 & 0.0176  & 0.5596 & -0.1159 & -0.0455 & 0.1964 & 0.3019
 & 0.0610 & 18.2 &  12.5 \\
 & E\pnote[b] & 5.531 &  5.758 &  7.975 &  254.0 &  1.0410 &
 1.4419 &  0.0149 &  0.5566 &  -0.1163 &  -0.0468 &  0.1935 &  0.3037
 &  0.0609 &  17.9 &  12.5 \\ 
 GdScO\3\ & T & 5.4987 & 5.7794 & 7.9861 & 253.79 & 1.0510 &
 1.4524 & 0.0191 &  0.5617 & -0.1222 & -0.0502 & 0.1941 &
 0.3036 & 0.0642 & 19.0 &  13.3 \\ 
 & E\pnote[b] & 5.481 &  5.745 &  7.929 &  249.7 &  1.0482 &
 1.4466 &  0.0163 &  0.5599 &  -0.1209 &  -0.0501 &  0.1912 &  0.3052
 &  0.0628 &  18.7 &  13.2 \\
 TbScO\3\ & T & 5.4764 & 5.7646 & 7.9735 & 251.72 & 1.0526 &
 1.4560 & 0.0198 &  0.5624 & -0.1251 & -0.0524 & 0.1932 & 0.3043
 & 0.0656 & 19.4 &  13.7 \\ 
 & E\pnote[k] & 5.4543 &  5.7233 &  7.9147 &  247.07 &  1.0493 &
 1.4511 &  0.0167 &  0.5603 &  -0.1239 &  -0.0545 &  0.1900 &  0.3054 &
 0.0643 &  19.1 &  13.7 \\
 DyScO\3\ & T & 5.4560 & 5.7501 & 7.9629 & 249.81 & 1.0539 &
 1.4595 & 0.0203 &  0.5630 & -0.1276 & -0.0545 & 0.1923 &
 0.3050 & 0.0669 & 19.7 &  14.1 \\ 
 & O\pnote[f] &  5.449 & 5.739 &  7.929 &  248.0 &  1.0532 &
 1.4551 & 0.019 &  0.562 &  -0.130 &  -0.057 &  0.190 &  0.307 &
 0.068 & 19.9 &  14.1 \\
 & E\pnote[b] & 5.443 &  5.717 &  7.901 &  245.9 &  1.0503 &
 1.4516 &  0.0174 &  0.5616 &  -0.1262 &  -0.0561 &  0.1886 &  0.3063
 &  0.0659 &  19.4 &  13.9 \\ 
LaYO\3\ & T & 5.9035 & 6.1225 & 8.5810 & 310.16 & 1.0371 & 1.4535 &
0.0173 & 0.5506 & -0.1284 & -0.0581 & 0.1948 & 0.3035 & 0.0689 &
18.9 & 13.6 \\ 
& E\pnote[l]& 5.890 &  6.086 &  8.511 &  305.1 &  1.0333 &
1.4450 &  &  &  &  &  &  &  & \\
 CaZrO\3\ & T & 5.5974 & 5.7875 & 8.0416 & 260.51 & 1.0340 & 1.4367 &
 0.0133 & 0.5506 & -0.1078 & -0.0413 & 0.1976 & 0.2999 & 0.0572
 & 16.6 & 11.6 \\ 
 & E\pnote[m] & 5.5831 &  5.7590 &  8.0070 &  257.45 &  1.0315 &
 1.4341 &  0.0122 &  0.5495 &  -0.1044 &  -0.0401 &  0.1976 &  0.3000 &
  0.0554 &  16.2 &  11.4 \\
 SrZrO\3\ & T & 5.8068 & 5.8602 & 8.2323 & 280.14 & 1.0092 &
 1.4177 & 0.0070  & 0.5311 & -0.0759 & -0.0195 & 0.2140 &
 0.2857 & 0.0401 & 11.6  & 7.7 \\ 
 & O\pnote[g] & 5.652 &  5.664 &  7.995 &  255.9 &  1.0021 &
 1.4145 &  0.007 &  0.534 &  -0.107 &  -0.036 &  0.199 &  0.301 &
 0.056 &  14.7 &  10.0 \\
 & E\pnote[c] & 5.7963 &  5.8171 &  8.2048 &  276.65 &  1.0036
 &  1.4155 &  0.0040 &  0.5242 &  -0.0687 &  -0.0133 &  0.2154 &
 0.2837 &  0.0363 &  10.4 &  7.2 \\
 SrHfO\3\ & T & 5.7552 & 5.7754 & 8.1365 & 270.45 & 1.0035 &
 1.4138 & 0.0052  & 0.5230 & -0.0660 & -0.0128 & 0.2209 &
 0.2792 & 0.0346 & 10.0  & 6.2 \\ 
 & O\pnote[h] & 5.6887 &  5.7016 &  8.0455 &  260.95 &  1.0023
 &  1.4143 &  0.006 &  0.528 &  -0.073 &  -0.016 &  0.2166 &  0.2834 &
 0.0385 &  10.8 &  7.0 \\
 & E\pnote[d] & 5.7516 &  5.7646 &  8.1344 &  269.70 &  1.0023 &
 1.4143 &  0.0040 &  0.5160 &  -0.0630 &  -0.0140 &  0.2189 &  0.2789 &
 0.0335 &  9.6 &  6.4 \\
LaHoO\3\ & T & 5.9135 & 6.1367 & 8.5859 & 311.58 & 1.0377 & 1.4519 &
0.0170  & 0.5508 & -0.1293 & -0.0587 & 0.1936 & 0.3044 & 0.0692 & 19.0 &
13.6 \\ 
& E\pnote[i] & 5.884 &  6.094 &  8.508 &  305.1 &  1.0357 &
1.4460 &   &   &   &   &  &   &   & \\
LaErO\3\ & T & 5.8971 & 6.1174 & 8.5509 & 308.48 & 1.0374 & 1.4500 &
0.0169 &
0.5509 & -0.1272 & -0.0567 & 0.1945 & 0.3036 & 0.0681 & 18.7 &
13.3 \\ 
& E\pnote[i] & 5.870 &  6.073 &  8.465 &  301.8 &  1.0346 &
1.4421 &   &   &   &   &  &   &   & \\
LaTmO\3\ & T & 5.8829 & 6.1002 & 8.5190 & 305.73 & 1.0369 & 1.4481 &
0.0167 &
0.5509 & -0.1252 & -0.0549 & 0.1951 & 0.3030 & 0.0670 & 18.5 & 13.1
\\ 
& E\pnote[i] & 5.859 &  6.047 &  8.453 &  299.5 &  1.0321 &
1.4427 &   &   &   &  &   &   &   & \\
LaYbO\3\ & T & 5.8692 & 6.0833 & 8.4890 & 303.09 & 1.0365 & 1.4464 &
0.0164 &
0.5509 & -0.1233 & -0.0532 & 0.1957 & 0.3024 & 0.0659 & 18.3 &
12.8 \\ 
& E\pnote[i] & 5.843 &  6.033 &  8.432 &  297.2 &  1.0325 &
1.4431 &   &   &   &  &   &   &   & \\
 LaLuO\3\ & T & 5.8579 & 6.0695 & 8.4646 & 300.95 & 1.0361 &
 1.4450 & 0.0162 & 0.5508 & -0.1217 & -0.0518 & 0.1962 & 0.3020 &
 0.0651 & 18.2 & 12.6 \\ 
 & E\pnote[e] & 5.8259 &  6.0218 &  8.3804 &  294.00 &  1.0336 &
 1.4385 &  0.0138 & 0.5507 &  -0.121 &  -0.056 &  0.193 &  0.307 &
 0.063 &  17.9 &  12.7 \\
CeTmO\3\ & T & 5.8520 & 6.0870 & 8.4984 & 302.72 & 1.0401 & 1.4522 &
0.0179 &
0.5528 & -0.1283 & -0.0575 & 0.1942 & 0.3036 & 0.0686 & 19.0 &
13.5 \\ 
& E\pnote[i] & 5.828 &  6.035 &  8.405 &  295.6 &  1.0355 &
1.4422 &   &   &   & &   &   &   & \\
CeYbO\3\ & T & 5.8381 & 6.0707 & 8.4671 & 300.09 & 1.0398 & 1.4503 &
0.0176
& 0.5529 & -0.1264 & -0.0558 & 0.1948 & 0.3031 & 0.0676 & 18.8 &
13.3 \\ 
& E\pnote[i] & 5.806 &  6.009 &  8.373 &  292.1 &  1.0350 &
1.4421 &   &   &   &   &  &   &   & \\
CeLuO\3\ & T & 5.8270 & 6.0578 & 8.4420 & 297.99 & 1.0396 & 1.4488 &
0.0174 &
0.5529 & -0.1249 & -0.0544 & 0.1952 & 0.3027 & 0.0667 & 18.7 &
13.1 \\ 
& E\pnote[i] & 5.793 &  5.997 &  8.344 &  289.9 &  1.0352 &
1.4404 &   &   &   &  &   &   &   & \\
PrYbO\3\ & T & 5.8085 & 6.0544 & 8.4481 & 297.10 & 1.0423 & 1.4544 &
0.0185 &
0.5542 & -0.1296 & -0.0585 & 0.1936 & 0.3039 & 0.0692 & 19.3 &
13.7 \\ 
& E\pnote[i] & 5.776 &  5.995 &  8.368 &  289.8 &  1.0379 &
1.4488 &   &   &   &  &   &   &   & \\
PrLuO\3\ & T & 5.7974 & 6.0424 & 8.4217 & 295.01 & 1.0423 & 1.4527 &
0.0183
& 0.5543 & -0.1281 & -0.0571 & 0.1936 & 0.3036 & 0.0683 & 19.1 &
13.5 \\ 
& E\pnote[i] & 5.768 &  5.991 &  8.340 &  288.2 &  1.0387 &
1.4459 &   &   &   &   &  &   &   & \\
NdLuO\3\ & T & 5.7699 & 6.0270 & 8.4062 & 292.32 & 1.0446 & 1.4569 &
0.0193 &
0.5555 & -0.1310 & -0.0596 & 0.1932 & 0.3042 & 0.0699 & 19.5 &
14.0 \\ 
& E\pnote[i] & 5.737 &  5.974 &  8.311 &  284.8 &  1.0413 &
1.4487 &   &   &   &  &   &   &   & \\
\end{tabular*}
\begin{tablenotes}[para]
\pitem[a]{Reference [\onlinecite{liferovich-exp}].}
\pitem[b]{Reference [\onlinecite{velickov-exp}].}
\pitem[c]{Reference [\onlinecite{kennedy-exp}].}
\pitem[d]{Reference [\onlinecite{kennedy2-exp}].}
\pitem[e]{Reference [\onlinecite{ito}].}
\pitem[f]{Reference [\onlinecite{delugas-th}].}
\pitem[g]{Reference [\onlinecite{vali-zr}].}
\pitem[h]{Reference [\onlinecite{vali-hf}].}
\pitem[i]{Reference [\onlinecite{berndt}].}
\pitem[j]{Reference [\onlinecite{gesing}].}
\pitem[k]{Reference [\onlinecite{velickov-tbsco}].}
\pitem[l]{Reference [\onlinecite{ruiz}].}
\pitem[m]{Reference [\onlinecite{cazro}].}
\end{tablenotes}
\end{ruledtabular} 
\end{threeparttable}
\end{table*}

\begin{table*}
  \caption{\label{tab:strbB} Structural parameters of the \labbo\ perovskites
    we considered, space group \rs. Angles are given in degrees,
    unit cell vectors in \AA\ and unit cell volumes in
    \AA$^3$. Monoclinic angle between $a$ and $c$ lattice vectors is
    denoted as $\alpha_{ac}$. {\it A}-site ions occupy the $4e$
    site with coordinate $(3/4, 1/2, 1/4)$. {\it B}-site ions
    (either Mg or Ca) occupy the $2a$ site and \bpr\ (Zr or Hf) occupy
    the $2d$ site. There are three non-equivalent positions for
    oxygens and they all occupy $4e$ sites. Coordinates of oxygen atoms
    in the cubic case are $(1/2,0,1/4)$, $(1/4,1/4,0)$, and
    $(3/4,1/4,0)$. Fitting parameters $\thR$, $\thM$, $\thM'$ and
    $d/d'$ are also given; see Sec.~\ref{sec:rigid} for the details.}
\begin{ruledtabular}
\begin{tabular*}{\textwidth}{
  l
  l r
  l r
  l D..{2.4} D..{2.4} D..{2.4}
  }
  & \multicolumn{2}{c}{Unit cell } & 
  \multicolumn{2}{c}{Model} & \multicolumn{4}{c}{Wyckoff  coordinates}  \\
  \cline{2-3}\cline{4-5}\cline{6-9} 
  \multirow{5}{*}{La$_2$MgZrO$_6$} &
  \quad $a$ & 5.6899 \quad &
  \quad $\thR$ & 15.7 \quad &
  & \mcc{$x$} & \mcc{$y$} & \mcc{$z$}
  \\
  \cline{7-9} 
  &
  \quad $b$ & 5.8169 \quad &
  \quad $\thM$ & 10.0 \quad &
  \quad La &  0.7634 & 0.4546 & 0.2489
  \\
  &
  \quad $c$ & 8.1274 \quad &
  \quad $\thM'$& 10.1 \quad & 
  \quad O$_{\rm I}$ & 0.6538 & 0.0305 & 0.2505
  \\
  &
  \quad $\alpha_{ac}$ & 90.301 \quad & 
  \quad $d/d'$ & 1.004 \quad &
  \quad O$_{\rm II}$ &  0.2411 & 0.2080 & 0.0523
  \\
  &
  \quad V & 269.00 \quad &
  & & 
  \quad O$_{\rm III}$ & 0.8429 & 0.2938 & -0.0499
  \\
\hline
  & \multicolumn{2}{c}{Unit cell } & 
  \multicolumn{2}{c}{Model} & \multicolumn{4}{c}{Wyckoff  coordinates}  \\
  \cline{2-3}\cline{4-5}\cline{6-9} 
  \multirow{5}{*}{La$_2$MgHfO$_6$} &
  \quad $a$ & 5.6669 \quad &
  \quad $\thR$ & 14.9 \quad &
  & \mcc{$x$} & \mcc{$y$} & \mcc{$z$}
  \\
  \cline{7-9} 
  &
  \quad $b$ & 5.7679 \quad &
  \quad $\thM$ & 9.3 \quad &
  \quad La & 0.7622 & 0.4578 & 0.2490
  \\
  &
  \quad $c$ & 8.0765 \quad &
  \quad $\thM'$& 9.5 \quad &
  \quad O$_{\rm I}$ & 0.6567 &  0.0270 & 0.2525 
  \\
  &
  \quad $\alpha_{ac}$ & 90.205 \quad & 
  \quad $d/d'$ & 1.018 \quad & 
  \quad O$_{\rm II}$ & 0.2430 &  0.2121 & 0.0498 
  \\
  &
  \quad V & 263.99 \quad &
  & &  
  \quad O$_{\rm III}$ & 0.8356 & 0.2928 & -0.0474 
  \\
\hline
  & \multicolumn{2}{c}{Unit cell } & 
  \multicolumn{2}{c}{Model} & \multicolumn{4}{c}{Wyckoff  coordinates}  \\
  \cline{2-3}\cline{4-5}\cline{6-9} 
  \multirow{5}{*}{La$_2$CaZrO$_6$} &
  \quad $a$ & 5.8188 \quad &
  \quad $\thR$ & 19.9 \quad &
  & \mcc{$x$} & \mcc{$y$} & \mcc{$z$}
  \\
  \cline{7-9} 
  &
  \quad $b$ & 6.0547 \quad &
  \quad $\thM$ & 12.0 \quad &
  \quad La & 0.7698 & 0.4469 &  0.2471 
  \\
  &
  \quad $c$ & 8.4245 \quad &
  \quad $\thM'$& 13.2 \quad &
  \quad O$_{\rm I}$ & 0.6134 & 0.0538 &  0.2649 
  \\
  &
  \quad $\alpha_{ac}$ &  90.182  \quad & 
  \quad $d/d'$ & 1.100 \quad & 
  \quad O$_{\rm II}$ & 0.2464 & 0.2123 & 0.0708 
  \\
  &
  \quad V & 296.80 \quad &
  & &  
  \quad O$_{\rm III}$ & 0.8500 &  0.3142 & -0.0588 
  \\
\hline
  & \multicolumn{2}{c}{Unit cell } & 
  \multicolumn{2}{c}{Model} & \multicolumn{4}{c}{Wyckoff  coordinates}  \\
  \cline{2-3}\cline{4-5}\cline{6-9} 
  \multirow{5}{*}{La$_2$CaHfO$_6$} &
  \quad $a$ & 5.7946 \quad &
  \quad $\thR$ & 19.3 \quad &
  & \mcc{$x$} & \mcc{$y$} & \mcc{$z$}
  \\
  \cline{7-9} 
  &
  \quad $b$ &  6.0211 \quad &
  \quad $\thM$ & 11.6 \quad &
  \quad La &  0.7687 &  0.4477 & 0.2474
  \\
  &
  \quad $c$ & 8.3680 \quad &
  \quad $\thM'$& 13.0 \quad &
  \quad O$_{\rm I}$ & 0.6155 &  0.0506 & 0.2668 
  \\
  &
  \quad $\alpha_{ac}$ &  90.100 \quad & 
  \quad $d/d'$ & 1.117 \quad & 
  \quad O$_{\rm II}$ & 0.2501 &  0.2150 & 0.0685 
  \\
  &
  \quad V &  291.96 \quad &
  & &  
  \quad O$_{\rm III}$ & 0.8447 & 0.3149 & -0.0568 
  \\
\end{tabular*}
\end{ruledtabular} 
\end{table*}

\subsection{Comparison with model of perfectly rigid octahedra}
\label{sec:rigid}

In \pbnm\ perovskites, a rigid rotation of the oxygen
octahedra by $\thR$ followed by another rigid rotation by $\thM$ (see
Fig.~\ref{fig:rot}) leads to Wyckoff parameters given by
\begin{align}
  x_2 & = - \frac{1}{2 \sqrt{2}} \tan \thR , \label{eq:fit1}\\
  y_2 & = - \frac{1}{2 \sqrt{2}} \sin \thR \tan \thM , \\
  x_3 & = \frac{1}{4} \left( 1 - \frac{\tan \thM}{\cos \thR} \right), \\
  y_3 & = \frac{1}{4} \left( 1 + \cos \thR \tan \thM \right) , \\
  z_3 & = \frac{1}{4 \sqrt{2}} \tan \thR,
\end{align}
Here we have denoted the Wyckoff coordinates of the oxygen atoms at the
$4c$ Wyckoff point with $x_2$ and $y_2$, while those of the
remaining oxygen atoms at the $8d$ point are denoted with $x_3$,
$y_3$, and $z_3$. The Wyckoff coordinates of the {\it A}-site ion at
the $4c$ point are denoted by $x_1$ and $y_1$, but these are left
unspecified in our rigid-octahedra model.
It also leads to orthorhombic lattice constants given by
\begin{align}
  a & = \sqrt[3]{\frac{V_0}{\sqrt{2}}} \cos \thR \cos \thM, \\
  b & = \sqrt[3]{\frac{V_0}{\sqrt{2}}} \cos \thM, \\
  c & = \sqrt[3]{2 V_0} \cos \thR \label{eq:fit8}
\end{align}
where $V_0$ is the volume the structure would have if the octahedra
were rotated rigidly back to $\thR=\thM=0$.

The Wyckoff parameters and unit-cell ratios from our calculations can
be well fitted by Eqs.~\ref{eq:fit1}-\ref{eq:fit8} (see
Table~\ref{tab:str} and Fig.~\ref{fig:str_data} for the values of
the fitted angles).
By far the largest discrepancy is found for
Wyckoff parameter $y_2$. For a typical system (e.g., LaScO\3) the
discrepancy between calculated and fitted $y_2$ values is about 0.016, or
50\% with respect to the difference from the cubic case. For the
remaining oxygen Wyckoff coefficients, the discrepancy averages about
0.003, or $\sim$5\%.

The rotation angles for the \labbo\ systems were obtained by fitting their
structural parameters to a slightly more complicated model of rigid
octahedra than the one given in Eqs.~\ref{eq:fit1}-\ref{eq:fit8}.  In
this model, we first change the relative sizes of {\it B}- and
\bpr-centered octahedra. The ratio of their linear sizes is denoted by
$d/d'$.  We then proceed with the rotation by an angle $\thR$
around the $[110]$ axis in the cubic frame. Finally, we perform
a rotation of the {\it B}-centered octahedra around $[001]$ by an angle
$\thM$, and of the \bpr-centered octahedra by an angle $\thM'$ around the same
axis.  The resulting fitted values of these parameters are given in
Table~\ref{tab:strbB}.
\subsection{Dielectric properties}
\label{sec:die}

In this section we discuss the dielectric properties of the
materials included in our study. \pbnm\ perovskites
are orthorhombic and thus have diagonal dielectric tensors, with
$\xe\ne\ye\ne\ze$ in general. In addition to reporting these
components, we also focus on analyzing the results in terms of
the three linear combinations
\begin{align}
\ae &= \displaystyle\frac{1}{3} \left( \xe+\ye+\ze \right), \label{eq:ae}\\
\1e &= \xe - \ye, \label{eq:1e} \\
\2e &= \ze - \displaystyle\frac{1}{2} \left( \xe+\ye \right), \label{eq:2e}
\end{align}
representing the average dielectric tensor, a measure of the
$x$-$y$ anisotropy, and a measure of $z$ anisotropy, respectively.
This choice of parameters was made to simplify the analysis of trends of
dielectric properties of these compounds.
The theoretical -- and where available, experimental -- results
for the dielectric-tensor components
are reported in Figure~\ref{fig:diel_data} and in Table~\ref{tab:die}.
The theoretical values are further decomposed in Table~\ref{tab:die}
into purely electronic or frozen-ion contributions $\epsilon^{\rm el}$
and lattice-mediated contributions $\epsilon^{\rm ion}$. We find that
the electronic contribution is roughly five times smaller than the
ionic one, is nearly isotropic, and does not show a dramatic variation
from one perovskite to another.  Thus, it is clear that the
lattice-mediated ionic contributions play by far the dominant role
in the observed dielectric tensors and their anisotropies.

Our calculations of the zone-center phonon frequencies as well as the
infrared activities for those modes that are infrared-active are given
in the supplementary material.~\cite{EPAPS_arxiv}

We now consider each of our chosen classes of \pbnm\ perovskites
in turn, orienting the presentation from the point of view of the
theoretical calculations, but mentioning the comparison with
experiment where appropriate.

\begin{figure}
  \includegraphics{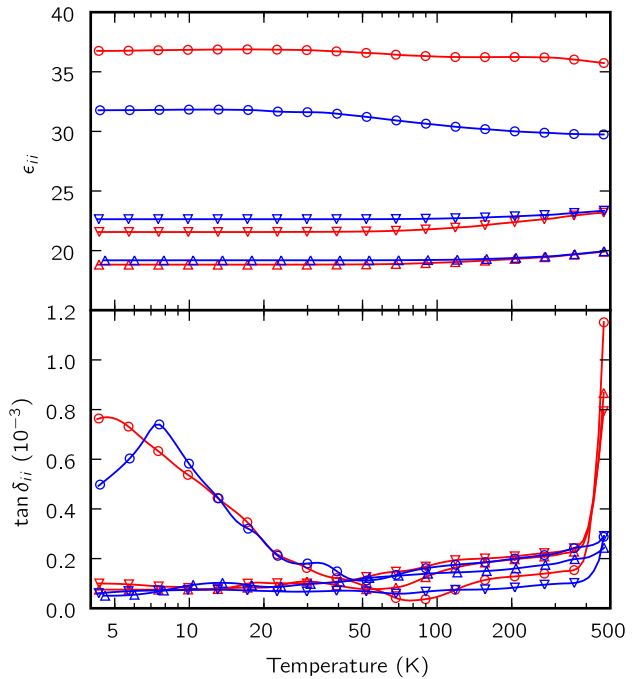}
  \caption{\label{fig:temp} Measured temperature dependence at 
    10~kHz of the
    dielectric tensor components (top panel) and dielectric loss (bottom
    panel) of GdScO\3\ (blue) and DyScO\3\ (red).  Downward-pointing
    triangles denote $\xe$ and $\tan \delta_{xx}$; upward-pointing
    triangles denote $\ye$ and $\tan \delta_{yy}$; circles
    denote $\ze$ and $\tan \delta_{zz}$.}
\end{figure}

\subsubsection{Rare-earth scandates}
\label{sec:resc_die}

All rare-earth scandates {\it A}ScO\3\ have rather similar values for
their isotropically-averaged dielectric constants, falling between about
$\ae=26$ and
$\ae=28$.  The $xx$ component for all these systems is larger than the
$yy$ component by about $\1e=4$. On the other hand, the $zz$ component
changes significantly from LaScO\3\ to DyScO\3. In LaScO\3 the average
of the $xx$ and $yy$ components is almost as large as the $zz$
component ($\2e$=-1), while in DyScO\3, the $zz$ component is larger
by about $\2e$=9 than the average of $xx$ and $yy$ components.

These results are in good agreement with experiment, especially for
$\ae$ and $\1e$. On the other hand, $\2e$ is consistently larger in
experiments by about 3-5, but the trend of increasing $\2e$ is present
in both theory and experiment.

As was mentioned earlier, rare-earth atoms heavier than Dy (i.e., Ho-Lu) and
Y itself do not form single-crystal scandates. Nevertheless, at least
some (YbScO\3~\cite{schubert} and LuScO\3~\cite{heeg-2}) can form
\pbnm\ perovskites in thin-film form.  In order to establish the
trends of the dielectric properties for these materials, we calculated the
dielectric tensors of LuScO\3\ and YScO\3. The dielectric tensor of
LuScO\3\ shows the continuation of the trend from LaScO\3\ to DyScO\3.
Both $xx$ and $yy$ components are slightly smaller than for
DyScO\3, their numerical values being 23.5 and 21.4 respectively.
On the other hand, the $zz$ component (44.8) is larger than for
DyScO\3\ (32.6) and for LaScO\3\ (27.4). YScO\3\ has
dielectric tensor components of 26.9, 23.0, and 37.7 for its
$xx$, $yy$, and $zz$ components, respectively.

The experimentally measured dependence of the dielectric tensor (and loss)
components 
on temperature is shown in Fig.~\ref{fig:temp} for two compounds,
GdScO\3\ and DyScO\3. In both cases we find that the dielectric tensor
properties 
do not change significantly with temperature over the examined
range ($4.2$-$470$~K). The $\ze$ component slightly
decreases with temperature while the $\xe$ and $\ye$ components show
the opposite
behavior. We expect that similar trends will be observed in all other 
rare-earth scandates.

As our calculated dielectric tensor is at 0~K and our measured
dielectric tensor (Table~\ref{tab:die} and \ref{tab:diebB})
is at room temperature, the absence of a
significant temperature dependence of the dielectric tensor is
important to our ability to make a meaningful comparison between the
calculated and measured coefficients.

Anomalies were observed in the dielectric loss along the
$z$ axis ($\tan \delta_{zz}$) 
for both DyScO\3\ and GdScO\3\ below 50~K. The origin of these is
unknown, but their presence suggests the possibility of a
low-temperature phase transition.

\subsubsection{Rare-earth yttrates}

We now consider the rare-earth yttrates, i.e.,
{\it A}YO\3 where {\it A} is one of the rare-earth
atoms.  These are similar to the rare-earth scandates, but with yttrium on
the {\it B} site instead of scandium.  Only one such compound, LaYO\3,
is known to form a perovskite,\cite{berndt} but others might form in
thin films via epitaxial stabilization.  We find that LaYO\3\ has a
larger $zz$ component than does LaScO\3. In LaYO\3\ the $zz$
component of the dielectric tensor is 38.0, while in LaScO\3\ it is
27.4. On the other hand, the $xx$ and $yy$ components are almost
unchanged with respect to LaScO\3. In LaYO\3\ the $xx$ component is
30.6 and the $yy$ component is 25.7. We also find that heavier
rare-earth atoms on the {\it A}-site tend to destabilize this \pbnm\
structure even further. For example, we find that DyYO\3\ in the
\pbnm\ structure has an
unstable mode at $i70$~cm$^{-1}$ that is IR-active along the $z$
direction.

\subsubsection{CaZrO\3, SrZrO\3, and SrHfO\3}
\label{sec:zrhf_die}

According to our calculations, SrZrO\3\ and SrHfO\3\ show large and
rather isotropic dielectric tensors. The average dielectric tensor $\ae$
is 40.9 and 32.8 in SrZrO\3\ and SrHfO\3, respectively.
Their $x$-$y$ anisotropies have
an opposite sign as compared to all of the other compounds we analyzed.
Unfortunately, because of twinning (see section \ref{sec:expElec}), we
could only measure an average dielectric constant for these systems,
and therefore we could not directly compare our full calculated
dielectric tensors with experiment. Still, if we make a comparison
between theory and experiment for the average dielectric tensor $\ae$,
the agreement is reasonable.

Our calculations suggest that the CaZrO\3\ compound, on the other hand, has
a very high value of the $z$-anisotropy of 28.1. Its $x$-$y$ anisotropy
of 1.1, on the other hand, is quite small. The average dielectric tensor
$\ae=43.6$ is the highest among the all compounds we considered, mostly
because of the very large $\ze$ component of the dielectric tensor.
Very similar results were also obtained in other theoretical
studies.~\cite{bennett,bennett-cazro}

\subsubsection{\labbo\ compounds}

The \labbo\ systems show a small, non-zero off-diagonal $\xze$
component, $-$0.4 for La$_2$MgZrO\3\ and 4 for La$_2$CaZrO\3. 
$\xze$ is allowed because the space group is reduced from
orthorhombic (\pbnm) to monoclinic (\rs) for these compounds.
Their isotropically-averaged
dielectric tensors are larger for systems containing
Ca than for those with Mg, and a bit larger for those with Zr than for
those with Hf.  Therefore, the  dielectric response in this class of
materials is largest for La$_2$CaZrO$_6$, with $\ae=28.5$, and smallest
for La$_2$MgHfO$_6$, with $\ae=23.6$. All computed dielectric tensor
components for these systems are given in Table~\ref{tab:diebB}.

\subsubsection{Rare-earth rare-earth perovskites}

The 11 rare-earth--rare-earth perovskites we considered show a bigger
variation in the isotropically-averaged dielectric constant $\ae$ than
do the rare-earth scandates (LaScO\3\ - DyScO\3).
The largest average dielectric constant among them is 32.9 in
LaHoO\3.  The largest component of a dielectric tensor is also
found in LaHoO\3, whose $\ze$ is 41.7.

The measure $\1e$ of $x$-$y$ anisotropy shows little variation among the
components in this series. The anisotropy is of the same sign as in
the rare-earth scandates.

Finally, the $z$ anisotropy $\2e$ once more shows a larger variation
than in the rare-earth scandates. This anisotropy is largest for
LaHoO\3\ and smallest for LaLuO\3.

\begin{table*}
  \caption{\label{tab:die} Dielectric parameters of the perovskites
    with space group \pbnm\ that
    we considered. Our calculated values are denoted by the letter T,
    our experimental values with E, and other theoretical data with O. 
    First all three non-zero dielectric constant tensor components are
    given. 
    Next, the average dielectric constant tensor, the $x$-$y$
    anisotropy, and $z$ anisotropy
    are given;
    see Eqs.~\ref{eq:ae}-\ref{eq:2e}. Finally, the electronic and ionic 
    contributions are given separately. }
\begin{ruledtabular}
\begin{tabular*}{\textwidth}{
  ll
  D..{2.2}D..{2.2}D..{2.2}
  D..{2.2}D..{2.2}D..{2.2}
  D..{1.2}D..{1.2}D..{1.2}
  D..{2.2}D..{2.2}D..{2.2}}
   & & \multicolumn{3}{c}{Dielectric tensor}  &
   \multicolumn{3}{c}{Reduced variables} &
   \multicolumn{3}{c}{Electronic part} & \multicolumn{3}{c}{Ionic
     part}\\
  \cline{3-5}\cline{6-8}\cline{9-11}\cline{12-14}
   & & \mcc{$\xe$} & \mcc{$\ye$} & \mcc{$\ze$} & 
  \mcc{$ \ae$} & \mcc{$\1e$} & \mcc{$\2e$} &
  \mcc{$\xe^{\rm el}$} & \mcc{$\ye^{\rm el}$} & \mcc{$\ze^{\rm el}$} &
  \mcc{$\xe^{\rm ion}$} & \mcc{$\ye^{\rm ion}$} &
  \mcc{$\ze^{\rm ion}$}  \\  
  \hline
   LaScO\3\ & T & 30.4 & 26.4 & 27.4 & 28.1 & 4.0 & -1.0 & 5.0
   & 5.0 & 4.8 & 25.3 & 21.5 & 22.6 \\
   PrScO\3\ & T & 28.6 & 24.2 & 26.2 & 26.3 & 4.4 & -0.1 & 5.0 
  & 5.0 & 4.8 & 23.5 & 19.2 & 21.5 \\
  & E\footnotemark[2] & 25.4 &  27.3 &  29.6  &  27.4 & -1.9 &  3.3 &
  &  &  &  &  & \\
   NdScO\3\ & T & 27.8 & 23.4 & 26.1 & 25.7 & 4.4 & 0.5 & 5.0 &
  4.9 & 4.7 & 22.8 & 18.5 & 21.3  \\
  & E\footnotemark[2] & 25.5 &  21.5 &  26.9 & 24.6 &  4.0 &  3.4 &
  &  &  &  &  &  \\
   SmScO\3\ & T & 27.2 & 22.9 & 27.3 & 25.8 & 4.3 & 2.3 & 4.9 &
  4.9 & 4.7 & 22.2 & 18.0 & 22.7 \\
  & E\footnotemark[2] & 23.1 &  19.9 &  29.0 & 24.0 &  3.2 &  7.5 &  &
  &  &  &  &  \\
   GdScO\3\ & T & 26.4 & 22.5 & 29.3  & 26.1 & 3.9 & 4.9 & 4.9
   & 4.8 & 4.6 & 21.6 & 17.7 & 24.7 \\
  & E\footnotemark[2] & 22.8 &  19.2 &  29.5 & 23.8 &  3.6 &  8.5 &  
  &  &  &  &  &  \\
   TbScO\3\ & T & 26.1 & 22.4 & 30.7 & 26.4 & 3.7 & 6.5 & 4.8 & 
  4.8 & 4.6 & 21.2 & 17.6 & 26.1 \\
   DyScO\3\ & T & 25.7 & 22.3 & 32.6 & 26.9 & 3.5 & 8.6 & 4.8 & 
  4.8 & 4.5 & 20.9 & 17.5 & 28.0  \\
  & O\footnotemark[1] & 24.1 & 21.2 & 27.7 & 24.3 & 2.9 & 5.1 & 4.9 & 
  4.9 & 4.7 & 19.2 & 16.3 & 23.0 \\
  & E\footnotemark[2] & 21.9 &  18.9 &  33.8 & 24.9 &  3.0 &  13.4 & 
  &  &  &  &  &  \\
LaYO\3\ & T & 30.6 & 25.7 & 38.0 &31.4 & 4.9 & 9.9 & 4.7 & 4.6
& 4.3 & 25.9 & 21.1 & 33.7 \\
  CaZrO\3\ & T & 34.8 & 33.7 & 62.4 & 43.6 & 1.1 & 28.1 & 4.6 & 4.7 &
  4.6 & 30.2 & 29.1 & 57.8 \\
  & E\footnotemark[6] &  &  &  &  30 &  & &  &  &  &  &  & \\
  SrZrO\3\ & T & 38.0 & 41.5 & 43.3 & 40.9 & -3.4 & 3.5 & 4.6 & 4.6 &
  4.6 & 33.4 & 36.9 & 38.7 \\
  & O\footnotemark[3] & 19.9 & 21.5 & 23.0 & 21.5 & -1.6 & 2.3 & 
  5.1 & 4.9 & 4.8 & 14.8 & 16.6 & 18.2 \\
  & E\footnotemark[2] &  &  &  &  32 &  & &  &  &  &  &  & \\
  SrHfO\3\ & T & 30.0 & 35.1 & 33.2 & 32.8 & -5.1 & 0.7 & 4.3 & 4.3 &
  4.3 & 25.7 & 30.9 & 29.0 \\
  & O\footnotemark[4] & 33.1 & 46.8 & 40.8 & 40.2 & -13.7 & 0.9
  & 4.4 & 4.3 & 4.3 & 28.7 & 42.4 & 36.5 \\
  & E\footnotemark[2] &  &  &  &  26.2 & &  &   &  &  &  &  & \\
LaHoO\3\ & T & 31.0 & 26.0 & 41.7 & 32.9 & 5.0 & 13.3 & 4.7 &
4.6 & 4.3 & 26.3 &21.4 & 37.4  \\
LaErO\3\ & T & 29.9 & 25.0 & 36.4 &30.4 & 4.9 & 9.0& 4.7 & 4.6 &
4.3 & 25.2 & 20.4 & 32.1  \\
LaTmO\3\ & T & 29.0 & 24.2 & 33.3 &28.8 & 4.8 & 6.7 & 4.6 &
4.6 & 4.3 & 24.4 & 19.7 & 29.0 \\
LaYbO\3\ & T & 28.3 & 23.6 & 30.9 &27.6 & 4.7 & 5.0 & 4.6 & 4.5
& 4.3 & 23.6 & 19.1 & 26.6  \\
  LaLuO\3\footnotemark[5] & T & 27.6 & 23.1 & 29.3 & 26.7 & 4.5 &
  3.9 & 4.6 & 
  4.5 & 4.3 & 23.0 & 18.6 & 25.0 \\
CeTmO\3\ & T & 27.9 & 23.5 & 34.9 & 28.8 & 4.4 & 9.2 & 4.6 &
4.6 & 4.3 & 23.3 & 18.9 & 30.6  \\
CeYbO\3\ & T & 27.2 & 22.8 & 32.1 & 27.4 & 4.4 & 7.0& 4.6 &
4.6 & 4.3 & 22.6 & 18.3 & 27.8  \\
CeLuO\3\ & T & 26.6 & 22.3 & 30.2 & 26.4 & 4.3 & 5.7& 4.6 & 4.5
& 4.3 & 22.0 & 17.8 & 25.9  \\
PrYbO\3\ & T & 26.6 & 22.7 & 34.7 &28.0 & 4.0 & 10.1& 4.6 &
4.5 & 4.3 & 22.0 & 18.2 & 30.4  \\
PrLuO\3\ & T & 26.1 & 22.0 & 32.2  &26.8 & 4.0 & 8.1 & 4.6 &
4.5 & 4.3 & 21.5 & 17.5 & 27.9\\
NdLuO\3\ & T & 25.3 & 21.7 & 34.5 &27.2 & 3.6 & 11.0& 4.6 &
4.5 & 4.3 & 20.8 & 17.2 & 30.2  \\
\end{tabular*}
\end{ruledtabular} 
\footnotetext[1]{Reference [\onlinecite{delugas-th}].}
\footnotetext[2]{This work.}
\footnotetext[3]{Reference [\onlinecite{vali-zr}].}
\footnotetext[4]{Reference [\onlinecite{vali-hf}].}
\footnotetext[5]{Experimental data in reference [\onlinecite{heeg-apl}].}
\footnotetext[6]{Reference [\onlinecite{cazro}].}
\end{table*}

\begin{table*}
  \caption{\label{tab:diebB} Dielectric parameters of the \labbo\
    perovskites we considered. First all four dielectric constant tensor
    components are given, followed by the average value of the
    dielectric constant tensor. Finally, the electronic and ionic
    contributions to dielectric constant tensor are given. $x$ axis is chosen
    along lattice vector $a$, $y$ along $b$, and $z$ close to $c$
    (with the small component along $a$ due to monoclinic cell).}
\begin{ruledtabular}
\begin{tabular*}{\textwidth}{
  l
  D..{2.2}D..{2.2}D..{2.2}D..{2.2}
  D..{2.2}
  D..{1.2}D..{1.2}D..{1.2}
  D..{2.2}
  D..{2.2}D..{2.2}D..{2.2}
  D..{2.2}}
  & \multicolumn{5}{c}{Dielectric tensor} & \multicolumn{4}{c}{Electronic
    part} & \multicolumn{4}{c}{Ionic part} \\ 
  \cline{2-6}\cline{7-10}\cline{11-14}
  & \mcc{$\xe$} & \mcc{$\ye$} & \mcc{$\ze$}
  & \mcc{$\xze$} & \mcc{$\ae$} &
  \mcc{$\xe^{\rm el}$} & \mcc{$\ye^{\rm el}$} & \mcc{$\ze^{\rm el}$} &  
  \mcc{$\xze^{\rm el}$} & \mcc{$\xe^{\rm ion}$} & \mcc{$\ye^{\rm ion}$} & 
  \mcc{$\ze^{\rm ion}$} & \mcc{$\xze^{\rm ion}$} \\  
  \hline
   La$_2$MgZrO$_6$ & 26.5 & 23.6 & 24.4 & -0.4 & 24.8 &
  4.8 & 4.6 & 4.5 & -0.06 &
  21.7 & 18.9 & 19.8 & -0.3 \\
   La$_2$MgHfO$_6$ & 24.9 & 22.9 & 22.9 & -0.3 & 23.6 &
  4.6 & 4.5 & 4.4 & -0.05 &
  20.3 & 18.4 & 18.5 & -0.3 \\
   La$_2$CaZrO$_6$ & 29.9 & 24.9 & 30.6 & 3.8 & 28.5 &
  4.7 & 4.6 & 4.4 & -0.06 &
  25.2 & 20.3 & 26.2 & 3.8 \\
   La$_2$CaHfO$_6$ & 27.4 & 23.0 & 26.6 & 2.6 & 25.7 &
  4.6 & 4.5 & 4.3 & -0.04 & 
  22.9 & 18.5 & 22.3 & 2.6 \\
\end{tabular*}
\end{ruledtabular} 
\end{table*}

\begin{figure*}[!t]
  \includegraphics{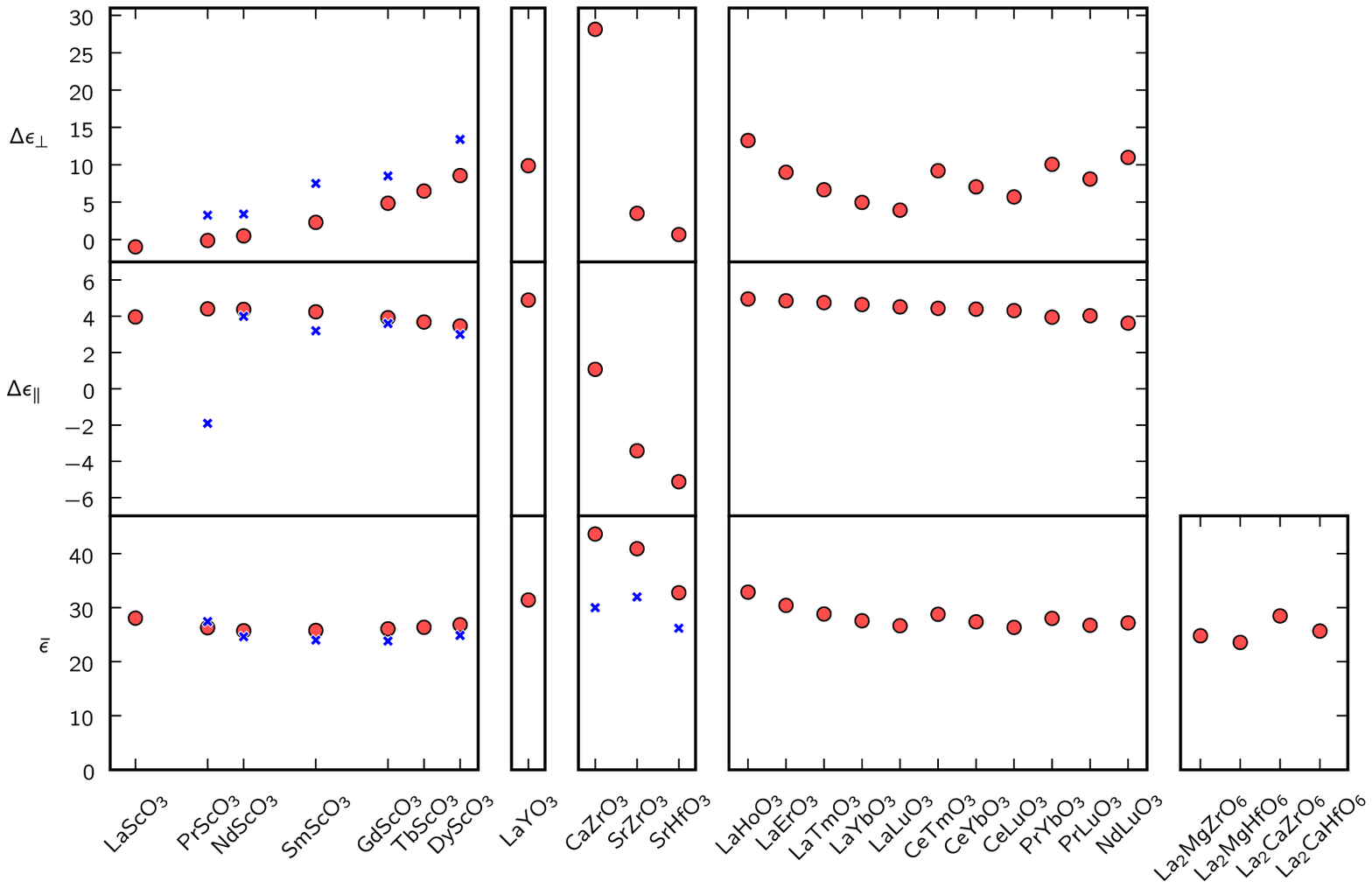}
  \caption{\label{fig:diel_data} Dielectric information for all
    systems we considered. Bottom pane shows average dielectric tensor
    $\ae$, middle pane shows $x$-$y$ anisotropy of the dielectric
    tensor, $\1e$, and $z$ anisotropy of the dielectric tensor $\2e$
    is given in top pane.  Calculated values are shown with red
    circles and experimental values with blue cross symbols, if
    available. See Table~\ref{tab:die} for numerical values.}
\end{figure*}

\subsection{Decomposition of the ionic contribution to the dielectric
  tensor}

\begin{figure*}
  \includegraphics{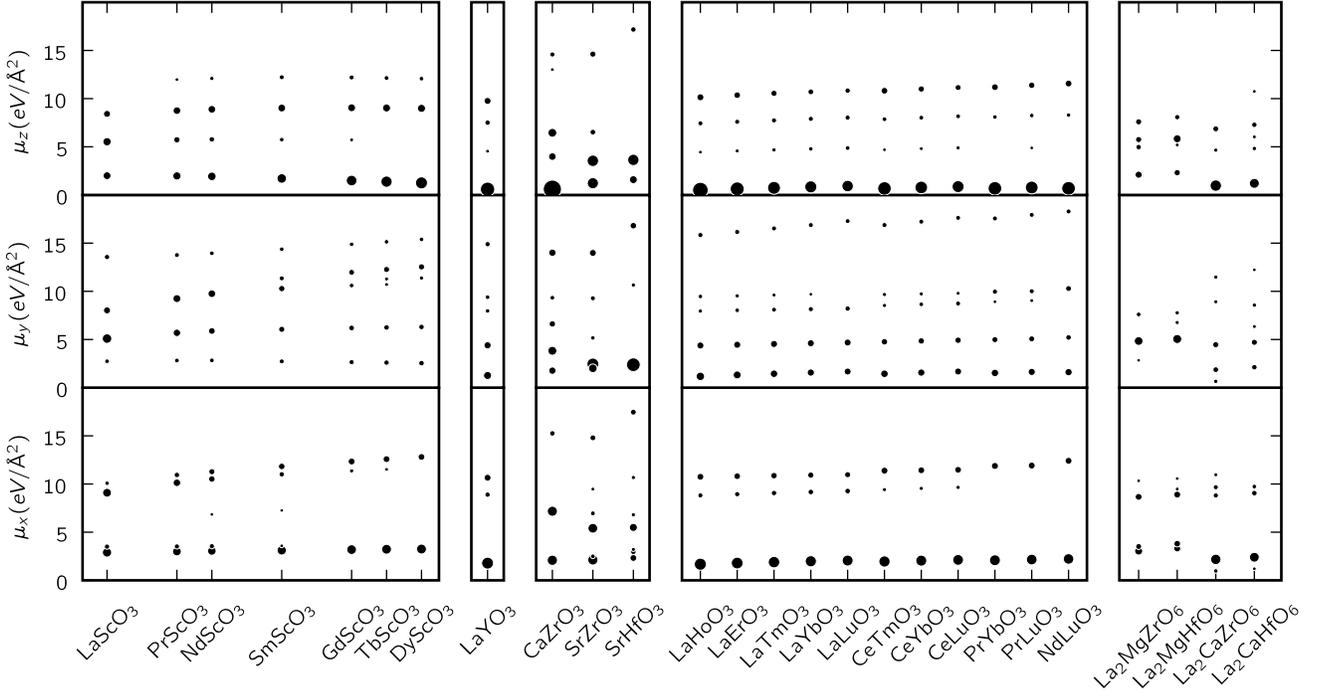}
  \caption{\label{fig:anis} Eigenvalues of a force constant matrix
    $\Phi_{i\alpha,j\beta}$ of modes that contribute to $\xe^{\rm
      ion}$ (first horizontal pane from bottom), $\ye^{\rm ion}$
    (second pane), or $\ze^{\rm ion}$ (third) by more than 1.5. For
    each mode the area of the circle is proportional to its
    contribution to $\epsilon_{ii}^{\rm ion}$. Eigenvalues are
    in eV/\AA$^2$.}
\end{figure*}

As already mentioned, the ionic contribution to the dielectric tensor
dominates in all of the systems we considered.  The expression
for the ionic contribution given in Eq.~(\ref{eq:eion}) provides a
decomposition into contributions coming from eigenmodes of the
force-constant matrix.
The \pbnm\ symmetry in perovskites, which is also approximately satisfied
in \labbo\ compounds, allows a given eigenmode to contribute only to a
single component ($\xe$, $\ye$, or $\ze$) of the dielectric tensor.
This decomposition is given in Fig.~\ref{fig:anis} for all three components.

In rare-earth scandates, all three directions are evidently very
different.  The $\xe$ component is dominated by a low-lying mode
whose contribution
is almost constant along the series (it contributes to $\xe$ by 9.6
for LaScO\3\ and 11.1 for DyScO\3). The $\ye$ component, on the other
hand, has sizable contributions coming from several modes. Finally, the
$\ze$ component comes mostly from a single low-lying mode. Unlike
for the $\xe$ component, the contribution from the mode responsible for
the $\ze$ component changes dramatically across the series, varying from
6.9 for LaScO\3\ to 16.3 for DyScO\3.  This explains the large
value of the $z$ anisotropy in DyScO\3\ as compared to LaScO\3 that
is visible in Fig.~\ref{fig:diel_data}.

A behavior similar to that of the rare-earth scandates is also observed
in LaYO\3\ and in the rare-earth rare-earth perovskites.
The SrZrO\3\ and SrHfO\3\ compounds show a quite similar behavior to each
other. The $\xe$ component has contributions coming from many modes,
the $\ye$ component is dominated by a single low-lying mode, and
$\ze$ is dominated by two low-lying modes. On the other hand, $\ze$
component in CaZrO\3\ shows very large contribution coming from a
single low-lying mode.
Finally, we note that the \labbo\
compounds containing Ca have stronger contributions to
$\xe$ and $\ze$ from low-lying modes than do those containing Mg.

\subsection{Compounds with \Irbar3c\ symmetry}
\label{sec:r3c}

At room temperature the ground state of BiFeO\3\ 
is ferroelectric with polar space group \rfe3c, the pattern of
octahedral rotations being \aaam\ in the Glazer notation. 
At higher temperature, however, BiFeO\3\ undergoes a phase transition in
which the ferroelectricity and the \aaam\ pattern of octahedral
rotations disappear simultaneously.\cite{arnold,haumont}
This observations led us to hypothesize that rotations of octahedra
around the pseudocubic $[111]$ axis, as in the \aaam\ pattern, tend to
be energetically compatible with the presence of a ferroelectric
distortion along the same axis. This would tend to suggest that
perovskites that adopt the centrosymmetric \rbar3c\ group, which
also exhibits the \aaam\ pattern of oxygen octahedra, might be
close to a ferroelectric instability leading to the lower-symmetry
\rfe3c\ space group, and thus that such compounds might have an
especially large component of the dielectric tensor along the
pseudocubic $[111]$ axis.

To test this hypothesis, we have carried out a series of calculations
on SrZrO\3\ and GdScO\3\ in which structural relaxation was allowed
while maintaining the \rbar3c\ symmetry. In both compounds we find
some IR-active phonon modes that either have very low or
imaginary frequency, indicating a near or actual instability.
In the case of SrZrO\3\ we find
a mode that is active along the $[111]$ pseudocubic
direction and has an extremely small frequency of only 6~cm$^{-1}$,
while for GdScO\3\ we find that the corresponding mode
is unstable with an imaginary frequency of i142~cm$^{-1}$.
These calculations show that imposing the \rbar3c\ structure on
SrZrO\3\ and GdScO\3\ make them nearly or actually ferroelectric,
thus confirming our hypothesis.

Incidentally, the observation that SrZrO\3\ is more likely than
GdScO\3\ to be stabilized in the \rbar3c\ structure is consistent with
the fact that
perovskite structures that prefer smaller rotation angles are more
likely to form \rbar3c\ than \pbnm\ structures, as discussed by
Woodward~\cite{woodward-oct}. We find that the rotational angles
for SrZrO\3\ in the \pbnm\ space group are $\thR=11.6^{\circ}$ and
$\thM=7.7^{\circ}$, while in GdScO\3\ they are substantially larger,
$\thR=19.0^{\circ}$ and $\thM=13.3^{\circ}$.
More directly, we also find that the ground-state energy of SrZrO\3\
having the \rbar3c\ structure is only higher by 33.6~meV per
formula unit than in the \pbnm\ structure. On the other
hand, in GdScO\3 the \rbar3c\ is higher in energy by a much larger
increment of 386~meV.

Finally, we note that LaAlO\3, NdAlO\3, and BaTbO\3\ may also be
of interest, as these all have the \rbar3c\ space-group symmetry
and should also be chemically stable on silicon.

\subsection{Correlation between structural and dielectric properties}
\label{sec:corr}

The heuristic observation about BiFeO\3\ mentioned in the previous
section (Sec.~\ref{sec:r3c}) led us to make a more detailed
analysis of the correlation between structural and dielectric anisotropies
in all five groups of \pbnm\ perovskites.
As can be seen from Fig.~\ref{fig:rot}, the presence of the octahedral
rotations breaks the symmetry among the three Cartesian directions in
the \pbnm\ perovskites. One would therefore naively expect that the
anisotropy in the dielectric tensor component should also be correlated
with the size of these rotation angles, but this is not what we
observe. For
example, LaScO\3\ and DyScO\3\ both have rather substantial
octahedral rotation angles ($\thR$ is 14.9$^{\circ}$ in LaScO\3\ and
19.7$^{\circ}$ in DyScO\3), but they have very different values of the
dielectric $z$ anisotropy ($\2e$ is $-$1.0 in LaScO\3\ and 8.6 in
DyScO\3). An even more extreme behavior can be seen in the case of
rare-earth perovskites between, e.g., LaHoO\3\ and LaLuO\3.

Thus, we find no simple correlation between the dielectric tensor
anisotropies and the values of the octahedral rotation angles.
Instead, we find a correlation between the dielectric tensor
anisotropies and the {\it mismatch} of the two rotation angles
$\thR$ and $\thM$, as we explain next.

While the \pbnm\ symmetry does not impose any relationship between
the two octahedral rotation angles $\thR$ and $\thM$, we find in
practice that all the compounds we studied obey the heuristic
relationship $\thR\simeq\sqrt{2}\thM$. This means that the oxygen
octahedra are rotated about the three Cartesian axes by almost the
same rotation angle, or equivalently, that the rotation axis is
nearly $\langle 111\rangle$. In the Glazer language, these
\pbnm\ perovskites having \aac\ rotations can be said to be
very close to an \aaap\ pattern.
We can measure the mismatch between the actual
\aaap\ and the hypothetical \aac\ rotation pattern by the
quantity $\thM - \thR/\sqrt{2}$, and it is this quantity that we
find to be strongly correlated with the dielectric anisotropy
$\2e$.

\begin{figure}[!h]
  \includegraphics{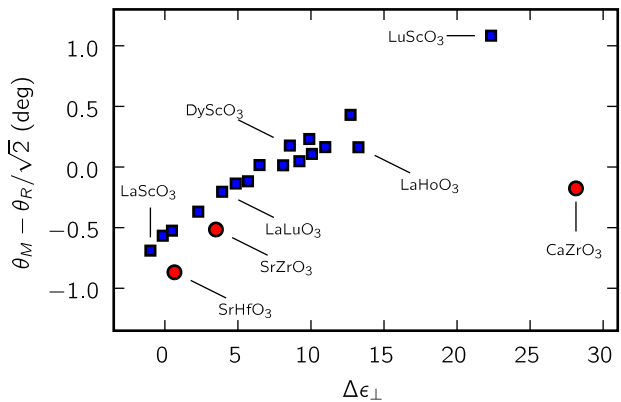}
  \caption{\label{fig:corr} Correlation between the dielectric tensor
    $z$-anisotropy ($\2e$) and the mismatch in the oxygen octahedra
    rotation angle. Only some perovskites are labeled. Perovskites
    where both {\it A} and {\it B} ions are III-valent are indicated
    with blue square symbols and those where {\it A} ions are
    II-valent and {\it B} are IV-valent are indicated with red circles.}
\end{figure}

This is shown in Fig.~\ref{fig:corr}, where $\2e$ is plotted versus
$\thM - \thR/\sqrt{2}$ for all of the compounds considered in this work.
It is apparent that the III-III--valent perovskites have a different
behavior than the II-IV--valent ones. Nevertheless, we conclude
that in both cases there is a strong correlation between the mismatch
angle and the dielectric tensor anisotropy. The sign of the correlation
is such that a deviation from the \aaap\ pattern having an increased
rotation angle around the $z$ axis gives a larger dielectric tensor
component along the $z$ axis, and thus a larger $z$ anisotropy $\2e$.

\subsection{Antisite substitutions}

Experimentally
the compositions of the perovskites that we have been describing up to
now by their nominal compositions, e.g., LaLuO\3, are in fact slightly
different from the compositions of the single crystals on which the
dielectric tensors were measured.  This is because our crystals are
grown at the congruently melting compositions, e.g.,
La$_{0.94}$Lu$_{1.06}$O$_3$,
which differ from the nominal compositions described up to now.  The
congruently melting compositions of all relevant \pbnm\ perovskites
studied have been found to be poor in the {\it A}-site cation and rich in
the {\it B}-site cation composition.~\cite{berkstresser,ovanesyan,gesing}

For this reason, we decided to carry out a theoretical analysis of
the effects of {\it B} atoms substituting at the {\it A} site on
the structural and dielectric properties of the material.  Detailed
calculations were done only for the case of LaLuO\3, but we expect
that similar trends will be observed in the remaining rare-earth
rare-earth perovskites as well as in the rare-earth scandates and
yttrates. Of course, other kinds of compositional disorder might
also be present, but such possibilities are not analyzed here.

\subsubsection{Analysis of antisite defects in LaLuO\3}

We studied Lu$_{\rm La}$ antisite defects in LaLuO\3\ using a
supercell approach.  Specifically, in order to model a situation
in which one of every 16 La atoms is substituted by Lu, which is
about a 6\% substitution, we constructed an 80-atom supercell
containing a single antisite defect.  The supercell is enlarged
with respect to the primitive 20-atom \pbnm\ primitive cell by
doubling along both the orthorhombic $a$ and $b$ lattice vectors.
The resulting stoichiometry is
\begin{align}
({\rm La}_{0.9375} {\rm Lu}_{0.0625}) {\rm Lu} {\rm O}_{3} 
{\rm \quad or \quad }
{\rm La}_{0.9375} {\rm Lu}_{1.0625} {\rm O}_{3}. \notag
\end{align}

The presence of the Lu$_{\rm La}$ antisite in this particular 80-atom
supercell reduces the crystal symmetry from orthorhombic \pbnm\ to
monoclinic $Pm$.
After full relaxation of the crystal structure in this
space group, we find that the $a$, $b$, and $c$ lattice vectors
are reduced by 0.3\%, 0.2\% ,and 0.1\%, respectively, while the
monoclinic angle between $a$ and $b$ lattice vectors of
$90.03^{\circ}$ deviates only very slightly from $90^{\circ}$.

The influence of the Lu$_{\rm La}$ substitution on the dielectric
properties is more complex. Evaluated in the same coordinate frame as
in the \pbnm\ unit cell, the $\xe$ and $\ye$ dielectric tensor components
remain almost unchanged, and the new $\epsilon_{xy}$ component allowed
by the monoclinic symmetry is quite small, only $2.1$. On the other
hand, the $\ze$ component is drastically altered by the presence
of Lu atom on the La site.  In fact, we find that the 80-atom
supercell is actually just barely unstable in the $Pm$ space group,
as indicated by the presence of a phonon mode with a very small
imaginary frequency of $i16$~cm$^{-1}$.~\cite{spoint}
The contribution of this phonon mode to the $\ze$ component (evaluated
in the unstable $Pm$ structure) is therefore negative, specifically,
$-33.6$.  Since this phonon frequency is so close to zero, we expect
that it would get renormalized to positive frequency at
room temperature. For this reason, we did not follow the structural
relaxation of our 80-atom supercell along the direction of the
unstable mode, and a realistic estimate of the dielectric
response of the system is difficult.  Nevertheless, we conclude that
Lu$_{\rm La}$ substitutions in LaLuO\3\ have the potential to
increase the $\ze$ dielectric tensor component substantially.

\subsubsection{Discussion}

As can be clearly seen in Fig.~\ref{fig:diel_data}, our calculated $z$
anisotropy ($\2e$) is consistently larger than the measured one
for all the rare-earth scandates for which we have experimental
measurements.  In view of the calculations reported for LaLuO\3 above, we
tentatively attribute this discrepancy to the generic tendency
of {\it B} atoms to substitute on the {\it A} site in these compounds.
This observation is consistent with the fact that smaller {\it B} ions
that substitute for larger {\it A} ions will reside in a relatively
larger cage, providing room to rattle and thereby contribute to an
enhanced dielectric response.

\section{Summary}
\label{sec:sum}

The main focus of this work has been the application of both
computational and experimental methods to study the structural and
dielectric properties of various \pbnm\ perovskites that have
potentially large dielectric tensor components and are chemically
stable on silicon
up to $\sim$1000~$^{\circ}$C.~\cite{ogale-oxide}
Such compounds might be good candidates for
future use as \hki\ dielectrics in microelectronics applications,
e.g., as a possible replacement of hafnia-based \hki\ dielectrics
currently used in the CMOS transistors in integrated circuits.

Of the compounds we have considered, CaZrO\3, SrZrO\3, LaHoO\3, and LaYO\3
appear to be especially promising. CaZrO\3\ has the largest calculated
average dielectric tensor ($\ae=43.6$) among the compounds we
considered, and SrZrO\3\ is a close second with $\ae=40.9$. The dielectric
tensor in CaZrO\3\ is very anisotropic, with its $\ze$ component
almost twice as large as $\xe$ or $\ye$, while on the other hand
SrZrO\3\ has an almost isotropic dielectric tensor. Unfortunately,
the full
dielectric tensors of these compounds have not yet been measured due to
lack of single crystals.

Of the rare-earth rare-earth \pbnm\ perovskites, only LaLuO\3\ has
had its dielectric tensor measured to date.  Our results on this
compound will be presented in detail elsewhere.~\cite{heeg-apl}
The theoretical calculations, however, indicate that other
compounds in this series should have even larger dielectric
tensor components, with LaHoO\3, having $\ae=32.9$, being the
most promising among these.  LaYO\3\ is expected to behave very
similar to LaHoO\3\ since Y and Ho have almost the same ionic
radii, so it may be promising as well ($\ae=31.4$). 
Thus, this series of
compounds clearly deserves additional scrutiny.

Of course, there are good reasons for preferring amorphous over
single-crystalline materials for such \hki\ applications.
Certainly the ability of amorphous SiO$_2$ to conform to the
substrate and to eliminate electrical traps played a central role
in its dominance as the gate dielectric 
of choice for 40 years for silicon-based metal-oxide-semiconductor
field-effect transistors.
The present
hafnia-based \hki\ dielectrics are amorphous or 
nanocrystalline.~\cite{muller}
For this reason, any eventual application of these materials for
\hki\ applications would presumably require the adoption of one of
two strategies.  The first is the possibility of growing crystalline
epitaxial oxides directly on silicon, which clearly would require
a very high level of control of interface chemistry and morphology
before it could become a practical solution.  The second is the
possibility that some of the compounds investigated here could be
synthesized in amorphous or nanocrystalline form.  We have not
investigated these issues here, nor have we tried to calculate
what (possibly very substantial) changes in the dielectric properties would
occur in the amorphous counterparts, as this would take us far
beyond the scope of the present study.  Nevertheless,
these are important questions for future investigations.

\begin{acknowledgments}
  The work of S.~C. and D.~V. was supported in part by NSF Grant
  DMR-0545198.
  The work of T.H. was supported by the Pennsylvania State University
  Materials Research Institute Nanofabrication Lab, the National
  Science Foundation Cooperative Agreement No. 0335765 and National
  Nanotechnology Infrastructure Network, with Cornell University.
  S.~T.~M. acknowledge support from NSF DMR-0602770.
  The research conducted by M.~D.~B. at the Center for Nanophase
  Materials Sciences, is sponsored at Oak Ridge National Laboratory by
  the Division of Scientific User Facilities, U.S. Department of
  Energy.
  D.~G.~S. would like to acknowledge support from the Semiconductor
  Research Corporation and Intel.
\end{acknowledgments}

\bibliography{pap}

\begin{thebibliography}{75}
\expandafter\ifx\csname natexlab\endcsname\relax\def\natexlab#1{#1}\fi
\expandafter\ifx\csname bibnamefont\endcsname\relax
  \def\bibnamefont#1{#1}\fi
\expandafter\ifx\csname bibfnamefont\endcsname\relax
  \def\bibfnamefont#1{#1}\fi
\expandafter\ifx\csname citenamefont\endcsname\relax
  \def\citenamefont#1{#1}\fi
\expandafter\ifx\csname url\endcsname\relax
  \def\url#1{\texttt{#1}}\fi
\expandafter\ifx\csname urlprefix\endcsname\relax\def\urlprefix{URL }\fi
\providecommand{\bibinfo}[2]{#2}
\providecommand{\eprint}[2][]{\url{#2}}

\bibitem[{\citenamefont{Moore}(2003)}]{moore}
\bibinfo{author}{\bibfnamefont{G.}~\bibnamefont{Moore}}, in
  \emph{\bibinfo{booktitle}{2003 IEEE International Solid-State Circuits
  Conference.}} (\bibinfo{address}{Piscataway, NJ, USA}, \bibinfo{year}{2003}),
  vol.~\bibinfo{volume}{1}, pp. \bibinfo{pages}{20--23}.

\bibitem[{\citenamefont{Schlom et~al.}(2005)\citenamefont{Schlom, Billman,
  Haeni, Lettieri, Tan, Held, V\"{o}lk, and Hubbard}}]{ogale-oxide}
\bibinfo{author}{\bibfnamefont{D.~G.} \bibnamefont{Schlom}},
  \bibinfo{author}{\bibfnamefont{C.~A.} \bibnamefont{Billman}},
  \bibinfo{author}{\bibfnamefont{J.~H.} \bibnamefont{Haeni}},
  \bibinfo{author}{\bibfnamefont{J.}~\bibnamefont{Lettieri}},
  \bibinfo{author}{\bibfnamefont{P.~H.} \bibnamefont{Tan}},
  \bibinfo{author}{\bibfnamefont{R.~R.~M.} \bibnamefont{Held}},
  \bibinfo{author}{\bibfnamefont{S.}~\bibnamefont{V\"{o}lk}}, \bibnamefont{and}
  \bibinfo{author}{\bibfnamefont{K.~J.} \bibnamefont{Hubbard}}, in
  \emph{\bibinfo{booktitle}{Thin Films and Heterostructures for Oxide
  Electronics}}, edited by \bibinfo{editor}{\bibfnamefont{S.~B.}
  \bibnamefont{Ogale}} (\bibinfo{publisher}{Springer}, \bibinfo{year}{2005}),
  pp. \bibinfo{pages}{31--78}.

\bibitem[{\citenamefont{Schlom et~al.}(2008)\citenamefont{Schlom, Guha, and
  Datta}}]{schlom-mrs}
\bibinfo{author}{\bibfnamefont{D.~G.} \bibnamefont{Schlom}},
  \bibinfo{author}{\bibfnamefont{S.}~\bibnamefont{Guha}}, \bibnamefont{and}
  \bibinfo{author}{\bibfnamefont{S.}~\bibnamefont{Datta}},
  \bibinfo{journal}{Mater. Res. Bull.} \textbf{\bibinfo{volume}{33}},
  \bibinfo{pages}{1017} (\bibinfo{year}{2008}).

\bibitem[{\citenamefont{Robertson}(2008)}]{robertson}
\bibinfo{author}{\bibfnamefont{J.}~\bibnamefont{Robertson}},
  \bibinfo{journal}{J. Appl. Phys.} \textbf{\bibinfo{volume}{104}},
  \bibinfo{eid}{124111} (\bibinfo{year}{2008}).

\bibitem[{\citenamefont{Mistry et~al.}(2007)\citenamefont{Mistry, Allen, Auth,
  Beattie, Bergstrom, Bost, Brazier, Buehler, Cappellani, Chau
  et~al.}}]{mistry}
\bibinfo{author}{\bibfnamefont{K.}~\bibnamefont{Mistry}},
  \bibinfo{author}{\bibfnamefont{C.}~\bibnamefont{Allen}},
  \bibinfo{author}{\bibfnamefont{C.}~\bibnamefont{Auth}},
  \bibinfo{author}{\bibfnamefont{B.}~\bibnamefont{Beattie}},
  \bibinfo{author}{\bibfnamefont{D.}~\bibnamefont{Bergstrom}},
  \bibinfo{author}{\bibfnamefont{M.}~\bibnamefont{Bost}},
  \bibinfo{author}{\bibfnamefont{M.}~\bibnamefont{Brazier}},
  \bibinfo{author}{\bibfnamefont{M.}~\bibnamefont{Buehler}},
  \bibinfo{author}{\bibfnamefont{A.}~\bibnamefont{Cappellani}},
  \bibinfo{author}{\bibfnamefont{R.}~\bibnamefont{Chau}}, \bibnamefont{et~al.},
  in \emph{\bibinfo{booktitle}{Electron Devices Meeting, 2007. IEDM 2007. IEEE
  International}} (\bibinfo{year}{2007}), pp. \bibinfo{pages}{247--250}.

\bibitem[{\citenamefont{Hicks et~al.}(2008)\citenamefont{Hicks, Bergstrom,
  Hattendorf, Jopling, Maiz, Pae, Prasad, and Wiedemer}}]{intel-highk-2}
\bibinfo{author}{\bibfnamefont{J.}~\bibnamefont{Hicks}},
  \bibinfo{author}{\bibfnamefont{D.}~\bibnamefont{Bergstrom}},
  \bibinfo{author}{\bibfnamefont{M.}~\bibnamefont{Hattendorf}},
  \bibinfo{author}{\bibfnamefont{J.}~\bibnamefont{Jopling}},
  \bibinfo{author}{\bibfnamefont{J.}~\bibnamefont{Maiz}},
  \bibinfo{author}{\bibfnamefont{S.}~\bibnamefont{Pae}},
  \bibinfo{author}{\bibfnamefont{C.}~\bibnamefont{Prasad}}, \bibnamefont{and}
  \bibinfo{author}{\bibfnamefont{J.}~\bibnamefont{Wiedemer}},
  \bibinfo{journal}{Intel Technol. J.} \textbf{\bibinfo{volume}{12}},
  \bibinfo{pages}{131} (\bibinfo{year}{2008}).

\bibitem[{\citenamefont{Scansen}(21 Jan. 2008)}]{scansen}
\bibinfo{author}{\bibfnamefont{D.}~\bibnamefont{Scansen}},
  \emph{\bibinfo{title}{Under the hood: 45 nm: What intel didn't tell you}}
  (\bibinfo{year}{21 Jan. 2008}),
  \bibinfo{note}{\href{http://www.techonline.com/product/underthehood/20591800%
4}{ www.techonline.com/product/underthehood/205918004}}.

\bibitem[{\citenamefont{Guha and Narayanan}(2007)}]{guha}
\bibinfo{author}{\bibfnamefont{S.}~\bibnamefont{Guha}} \bibnamefont{and}
  \bibinfo{author}{\bibfnamefont{V.}~\bibnamefont{Narayanan}},
  \bibinfo{journal}{Phys. Rev. Lett.} \textbf{\bibinfo{volume}{98}},
  \bibinfo{pages}{196101} (\bibinfo{year}{2007}).

\bibitem[{\citenamefont{Christen et~al.}(2006)\citenamefont{Christen,
  G.~E.~Jellison, Ohkubo, Huang, Reeves, Cicerrella, Freeouf, Jia, and
  Schlom}}]{christen}
\bibinfo{author}{\bibfnamefont{H.~M.} \bibnamefont{Christen}},
  \bibinfo{author}{\bibfnamefont{J.}~\bibnamefont{G.~E.~Jellison}},
  \bibinfo{author}{\bibfnamefont{I.}~\bibnamefont{Ohkubo}},
  \bibinfo{author}{\bibfnamefont{S.}~\bibnamefont{Huang}},
  \bibinfo{author}{\bibfnamefont{M.~E.} \bibnamefont{Reeves}},
  \bibinfo{author}{\bibfnamefont{E.}~\bibnamefont{Cicerrella}},
  \bibinfo{author}{\bibfnamefont{J.~L.} \bibnamefont{Freeouf}},
  \bibinfo{author}{\bibfnamefont{Y.}~\bibnamefont{Jia}}, \bibnamefont{and}
  \bibinfo{author}{\bibfnamefont{D.~G.} \bibnamefont{Schlom}},
  \bibinfo{journal}{Appl. Phys. Lett.} \textbf{\bibinfo{volume}{88}},
  \bibinfo{eid}{262906} (\bibinfo{year}{2006}).

\bibitem[{\citenamefont{Afanas'ev et~al.}(2004)\citenamefont{Afanas'ev,
  Stesmans, Zhao, Caymax, Heeg, Schubert, Jia, Schlom, and
  Lucovsky}}]{afanasev-1}
\bibinfo{author}{\bibfnamefont{V.~V.} \bibnamefont{Afanas'ev}},
  \bibinfo{author}{\bibfnamefont{A.}~\bibnamefont{Stesmans}},
  \bibinfo{author}{\bibfnamefont{C.}~\bibnamefont{Zhao}},
  \bibinfo{author}{\bibfnamefont{M.}~\bibnamefont{Caymax}},
  \bibinfo{author}{\bibfnamefont{T.}~\bibnamefont{Heeg}},
  \bibinfo{author}{\bibfnamefont{J.}~\bibnamefont{Schubert}},
  \bibinfo{author}{\bibfnamefont{Y.}~\bibnamefont{Jia}},
  \bibinfo{author}{\bibfnamefont{D.~G.} \bibnamefont{Schlom}},
  \bibnamefont{and} \bibinfo{author}{\bibfnamefont{G.}~\bibnamefont{Lucovsky}},
  \bibinfo{journal}{Appl. Phys. Lett.} \textbf{\bibinfo{volume}{85}},
  \bibinfo{pages}{5917} (\bibinfo{year}{2004}).

\bibitem[{\citenamefont{Zhao et~al.}(2005)\citenamefont{Zhao, Witters, Brijs,
  Bender, Richard, Caymax, Heeg, Schubert, Afanas'ev, Stesmans et~al.}}]{zhao}
\bibinfo{author}{\bibfnamefont{C.}~\bibnamefont{Zhao}},
  \bibinfo{author}{\bibfnamefont{T.}~\bibnamefont{Witters}},
  \bibinfo{author}{\bibfnamefont{B.}~\bibnamefont{Brijs}},
  \bibinfo{author}{\bibfnamefont{H.}~\bibnamefont{Bender}},
  \bibinfo{author}{\bibfnamefont{O.}~\bibnamefont{Richard}},
  \bibinfo{author}{\bibfnamefont{M.}~\bibnamefont{Caymax}},
  \bibinfo{author}{\bibfnamefont{T.}~\bibnamefont{Heeg}},
  \bibinfo{author}{\bibfnamefont{J.}~\bibnamefont{Schubert}},
  \bibinfo{author}{\bibfnamefont{V.~V.} \bibnamefont{Afanas'ev}},
  \bibinfo{author}{\bibfnamefont{A.}~\bibnamefont{Stesmans}},
  \bibnamefont{et~al.}, \bibinfo{journal}{Appl. Phys. Lett.}
  \textbf{\bibinfo{volume}{86}}, \bibinfo{eid}{132903} (\bibinfo{year}{2005}).

\bibitem[{\citenamefont{Heeg et~al.}(2006)\citenamefont{Heeg, Schubert, Buchal,
  Cicerrella, Freeouf, Tian, Jia, and Schlom}}]{heeg}
\bibinfo{author}{\bibfnamefont{T.}~\bibnamefont{Heeg}},
  \bibinfo{author}{\bibfnamefont{J.}~\bibnamefont{Schubert}},
  \bibinfo{author}{\bibfnamefont{C.}~\bibnamefont{Buchal}},
  \bibinfo{author}{\bibfnamefont{E.}~\bibnamefont{Cicerrella}},
  \bibinfo{author}{\bibfnamefont{J.}~\bibnamefont{Freeouf}},
  \bibinfo{author}{\bibfnamefont{W.}~\bibnamefont{Tian}},
  \bibinfo{author}{\bibfnamefont{Y.}~\bibnamefont{Jia}}, \bibnamefont{and}
  \bibinfo{author}{\bibfnamefont{D.}~\bibnamefont{Schlom}},
  \bibinfo{journal}{Appl. Phys. A} \textbf{\bibinfo{volume}{83}},
  \bibinfo{pages}{103} (\bibinfo{year}{2006}).

\bibitem[{\citenamefont{Edge et~al.}(2006{\natexlab{a}})\citenamefont{Edge,
  Schlom, Stemmer, Lucovsky, and Luning}}]{edge-1}
\bibinfo{author}{\bibfnamefont{L.}~\bibnamefont{Edge}},
  \bibinfo{author}{\bibfnamefont{D.}~\bibnamefont{Schlom}},
  \bibinfo{author}{\bibfnamefont{S.}~\bibnamefont{Stemmer}},
  \bibinfo{author}{\bibfnamefont{G.}~\bibnamefont{Lucovsky}}, \bibnamefont{and}
  \bibinfo{author}{\bibfnamefont{J.}~\bibnamefont{Luning}},
  \bibinfo{journal}{Radiat. Phys. Chem.} \textbf{\bibinfo{volume}{75}},
  \bibinfo{pages}{1608 } (\bibinfo{year}{2006}{\natexlab{a}}).

\bibitem[{\citenamefont{Cicerrella et~al.}(2005)\citenamefont{Cicerrella,
  Freeouf, Edge, Schlom, Heeg, Schubert, and Chambers}}]{cicerrella}
\bibinfo{author}{\bibfnamefont{E.}~\bibnamefont{Cicerrella}},
  \bibinfo{author}{\bibfnamefont{J.~L.} \bibnamefont{Freeouf}},
  \bibinfo{author}{\bibfnamefont{L.~F.} \bibnamefont{Edge}},
  \bibinfo{author}{\bibfnamefont{D.~G.} \bibnamefont{Schlom}},
  \bibinfo{author}{\bibfnamefont{T.}~\bibnamefont{Heeg}},
  \bibinfo{author}{\bibfnamefont{J.}~\bibnamefont{Schubert}}, \bibnamefont{and}
  \bibinfo{author}{\bibfnamefont{S.~A.} \bibnamefont{Chambers}},
  \bibinfo{journal}{J. Vac. Sci. Technol., A} \textbf{\bibinfo{volume}{23}},
  \bibinfo{pages}{1676} (\bibinfo{year}{2005}).

\bibitem[{\citenamefont{Afanas'ev et~al.}(2006)\citenamefont{Afanas'ev,
  Stesmans, Edge, Schlom, Heeg, and Schubert}}]{afanasev-2}
\bibinfo{author}{\bibfnamefont{V.~V.} \bibnamefont{Afanas'ev}},
  \bibinfo{author}{\bibfnamefont{A.}~\bibnamefont{Stesmans}},
  \bibinfo{author}{\bibfnamefont{L.~F.} \bibnamefont{Edge}},
  \bibinfo{author}{\bibfnamefont{D.~G.} \bibnamefont{Schlom}},
  \bibinfo{author}{\bibfnamefont{T.}~\bibnamefont{Heeg}}, \bibnamefont{and}
  \bibinfo{author}{\bibfnamefont{J.}~\bibnamefont{Schubert}},
  \bibinfo{journal}{Appl. Phys. Lett.} \textbf{\bibinfo{volume}{88}},
  \bibinfo{eid}{032104} (\bibinfo{year}{2006}).

\bibitem[{\citenamefont{Edge et~al.}(2006{\natexlab{b}})\citenamefont{Edge,
  Schlom, Rivillon, Chabal, Agustin, Stemmer, Lee, Kim, Craft, Maria
  et~al.}}]{edge-2}
\bibinfo{author}{\bibfnamefont{L.~F.} \bibnamefont{Edge}},
  \bibinfo{author}{\bibfnamefont{D.~G.} \bibnamefont{Schlom}},
  \bibinfo{author}{\bibfnamefont{S.}~\bibnamefont{Rivillon}},
  \bibinfo{author}{\bibfnamefont{Y.~J.} \bibnamefont{Chabal}},
  \bibinfo{author}{\bibfnamefont{M.~P.} \bibnamefont{Agustin}},
  \bibinfo{author}{\bibfnamefont{S.}~\bibnamefont{Stemmer}},
  \bibinfo{author}{\bibfnamefont{T.}~\bibnamefont{Lee}},
  \bibinfo{author}{\bibfnamefont{M.~J.} \bibnamefont{Kim}},
  \bibinfo{author}{\bibfnamefont{H.~S.} \bibnamefont{Craft}},
  \bibinfo{author}{\bibfnamefont{J.-P.} \bibnamefont{Maria}},
  \bibnamefont{et~al.}, \bibinfo{journal}{Appl. Phys. Lett.}
  \textbf{\bibinfo{volume}{89}}, \bibinfo{eid}{062902}
  (\bibinfo{year}{2006}{\natexlab{b}}).

\bibitem[{\citenamefont{Sivasubramani et~al.}(2006)\citenamefont{Sivasubramani,
  Lee, Kim, Kim, Gnade, Wallace, Edge, Schlom, Stevie, Garcia
  et~al.}}]{sivasubramani}
\bibinfo{author}{\bibfnamefont{P.}~\bibnamefont{Sivasubramani}},
  \bibinfo{author}{\bibfnamefont{T.~H.} \bibnamefont{Lee}},
  \bibinfo{author}{\bibfnamefont{M.~J.} \bibnamefont{Kim}},
  \bibinfo{author}{\bibfnamefont{J.}~\bibnamefont{Kim}},
  \bibinfo{author}{\bibfnamefont{B.~E.} \bibnamefont{Gnade}},
  \bibinfo{author}{\bibfnamefont{R.~M.} \bibnamefont{Wallace}},
  \bibinfo{author}{\bibfnamefont{L.~F.} \bibnamefont{Edge}},
  \bibinfo{author}{\bibfnamefont{D.~G.} \bibnamefont{Schlom}},
  \bibinfo{author}{\bibfnamefont{F.~A.} \bibnamefont{Stevie}},
  \bibinfo{author}{\bibfnamefont{R.}~\bibnamefont{Garcia}},
  \bibnamefont{et~al.}, \bibinfo{journal}{Appl. Phys. Lett.}
  \textbf{\bibinfo{volume}{89}}, \bibinfo{eid}{242907} (\bibinfo{year}{2006}).

\bibitem[{\citenamefont{Lopes et~al.}(2007)\citenamefont{Lopes, Roeckerath,
  Heeg, Littmark, Schubert, Mantl, Jia, and Schlom}}]{lopes}
\bibinfo{author}{\bibfnamefont{J.}~\bibnamefont{Lopes}},
  \bibinfo{author}{\bibfnamefont{M.}~\bibnamefont{Roeckerath}},
  \bibinfo{author}{\bibfnamefont{T.}~\bibnamefont{Heeg}},
  \bibinfo{author}{\bibfnamefont{U.}~\bibnamefont{Littmark}},
  \bibinfo{author}{\bibfnamefont{J.}~\bibnamefont{Schubert}},
  \bibinfo{author}{\bibfnamefont{S.}~\bibnamefont{Mantl}},
  \bibinfo{author}{\bibfnamefont{Y.}~\bibnamefont{Jia}}, \bibnamefont{and}
  \bibinfo{author}{\bibfnamefont{D.}~\bibnamefont{Schlom}},
  \bibinfo{journal}{Microelectron. Eng.} \textbf{\bibinfo{volume}{84}},
  \bibinfo{pages}{1890 } (\bibinfo{year}{2007}).

\bibitem[{\citenamefont{Wang et~al.}(2007)\citenamefont{Wang, He, Ma, Edge, and
  Schlom}}]{wang}
\bibinfo{author}{\bibfnamefont{M.}~\bibnamefont{Wang}},
  \bibinfo{author}{\bibfnamefont{W.}~\bibnamefont{He}},
  \bibinfo{author}{\bibfnamefont{T.~P.} \bibnamefont{Ma}},
  \bibinfo{author}{\bibfnamefont{L.~F.} \bibnamefont{Edge}}, \bibnamefont{and}
  \bibinfo{author}{\bibfnamefont{D.~G.} \bibnamefont{Schlom}},
  \bibinfo{journal}{Appl. Phys. Lett.} \textbf{\bibinfo{volume}{90}},
  \bibinfo{eid}{053502} (\bibinfo{year}{2007}).

\bibitem[{\citenamefont{Edge et~al.}(2008)\citenamefont{Edge, Tian,
  Vaithyanathan, Heeg, Schlom, Klenov, Stemmer, Wang, and Kim}}]{edge-3}
\bibinfo{author}{\bibfnamefont{L.}~\bibnamefont{Edge}},
  \bibinfo{author}{\bibfnamefont{W.}~\bibnamefont{Tian}},
  \bibinfo{author}{\bibfnamefont{V.}~\bibnamefont{Vaithyanathan}},
  \bibinfo{author}{\bibfnamefont{T.}~\bibnamefont{Heeg}},
  \bibinfo{author}{\bibfnamefont{D.}~\bibnamefont{Schlom}},
  \bibinfo{author}{\bibfnamefont{D.}~\bibnamefont{Klenov}},
  \bibinfo{author}{\bibfnamefont{S.}~\bibnamefont{Stemmer}},
  \bibinfo{author}{\bibfnamefont{J.}~\bibnamefont{Wang}}, \bibnamefont{and}
  \bibinfo{author}{\bibfnamefont{M.}~\bibnamefont{Kim}}, in
  \emph{\bibinfo{booktitle}{ECS Trans.}} (\bibinfo{address}{Honolulu, HI,
  United states}, \bibinfo{year}{2008}), vol.~\bibinfo{volume}{16}, pp.
  \bibinfo{pages}{213 -- 227}.

\bibitem[{\citenamefont{\"{O}zben et~al.}(2008)\citenamefont{\"{O}zben, Lopes,
  Roeckerath, Lenk, Holl\"{a}nder, Jia, Schlom, Schubert, and Mantl}}]{ozben}
\bibinfo{author}{\bibfnamefont{E.~D.} \bibnamefont{\"{O}zben}},
  \bibinfo{author}{\bibfnamefont{J.~M.~J.} \bibnamefont{Lopes}},
  \bibinfo{author}{\bibfnamefont{M.}~\bibnamefont{Roeckerath}},
  \bibinfo{author}{\bibfnamefont{S.}~\bibnamefont{Lenk}},
  \bibinfo{author}{\bibfnamefont{B.}~\bibnamefont{Holl\"{a}nder}},
  \bibinfo{author}{\bibfnamefont{Y.}~\bibnamefont{Jia}},
  \bibinfo{author}{\bibfnamefont{D.~G.} \bibnamefont{Schlom}},
  \bibinfo{author}{\bibfnamefont{J.}~\bibnamefont{Schubert}}, \bibnamefont{and}
  \bibinfo{author}{\bibfnamefont{S.}~\bibnamefont{Mantl}},
  \bibinfo{journal}{Appl. Phys. Lett.} \textbf{\bibinfo{volume}{93}},
  \bibinfo{eid}{052902} (\bibinfo{year}{2008}).

\bibitem[{\citenamefont{Roeckerath et~al.}(2009)\citenamefont{Roeckerath,
  Lopes, Durgun~Ozben, Sandow, Lenk, Heeg, Schubert, and Mantl}}]{roeckerath}
\bibinfo{author}{\bibfnamefont{M.}~\bibnamefont{Roeckerath}},
  \bibinfo{author}{\bibfnamefont{J.}~\bibnamefont{Lopes}},
  \bibinfo{author}{\bibfnamefont{E.}~\bibnamefont{Durgun~Ozben}},
  \bibinfo{author}{\bibfnamefont{C.}~\bibnamefont{Sandow}},
  \bibinfo{author}{\bibfnamefont{S.}~\bibnamefont{Lenk}},
  \bibinfo{author}{\bibfnamefont{T.}~\bibnamefont{Heeg}},
  \bibinfo{author}{\bibfnamefont{J.}~\bibnamefont{Schubert}}, \bibnamefont{and}
  \bibinfo{author}{\bibfnamefont{S.}~\bibnamefont{Mantl}},
  \bibinfo{journal}{Appl. Phys. A} \textbf{\bibinfo{volume}{94}},
  \bibinfo{pages}{521 } (\bibinfo{year}{2009}).

\bibitem[{\citenamefont{Wagner et~al.}(2006)\citenamefont{Wagner, Heeg,
  Schubert, Zhao, Richard, Caymax, Afanas'ev, and Mantl}}]{wagner}
\bibinfo{author}{\bibfnamefont{M.}~\bibnamefont{Wagner}},
  \bibinfo{author}{\bibfnamefont{T.}~\bibnamefont{Heeg}},
  \bibinfo{author}{\bibfnamefont{J.}~\bibnamefont{Schubert}},
  \bibinfo{author}{\bibfnamefont{C.}~\bibnamefont{Zhao}},
  \bibinfo{author}{\bibfnamefont{O.}~\bibnamefont{Richard}},
  \bibinfo{author}{\bibfnamefont{M.}~\bibnamefont{Caymax}},
  \bibinfo{author}{\bibfnamefont{V.}~\bibnamefont{Afanas'ev}},
  \bibnamefont{and} \bibinfo{author}{\bibfnamefont{S.}~\bibnamefont{Mantl}},
  \bibinfo{journal}{Solid-State Electron.} \textbf{\bibinfo{volume}{50}},
  \bibinfo{pages}{58 } (\bibinfo{year}{2006}).

\bibitem[{\citenamefont{Myllymaki et~al.}(Sept. 2007)\citenamefont{Myllymaki,
  Roeckerath, Putkonen, Lenk, Schubert, Niinisto, and Mantl}}]{myllymaki}
\bibinfo{author}{\bibfnamefont{P.}~\bibnamefont{Myllymaki}},
  \bibinfo{author}{\bibfnamefont{M.}~\bibnamefont{Roeckerath}},
  \bibinfo{author}{\bibfnamefont{M.}~\bibnamefont{Putkonen}},
  \bibinfo{author}{\bibfnamefont{S.}~\bibnamefont{Lenk}},
  \bibinfo{author}{\bibfnamefont{J.}~\bibnamefont{Schubert}},
  \bibinfo{author}{\bibfnamefont{L.}~\bibnamefont{Niinisto}}, \bibnamefont{and}
  \bibinfo{author}{\bibfnamefont{S.}~\bibnamefont{Mantl}},
  \bibinfo{journal}{Appl. Phys. A} \textbf{\bibinfo{volume}{A88}},
  \bibinfo{pages}{633 } (\bibinfo{year}{Sept. 2007}).

\bibitem[{\citenamefont{Kim et~al.}(2006)\citenamefont{Kim, Farmer, Lehn, Rao,
  and Gordon}}]{kim}
\bibinfo{author}{\bibfnamefont{K.~H.} \bibnamefont{Kim}},
  \bibinfo{author}{\bibfnamefont{D.~B.} \bibnamefont{Farmer}},
  \bibinfo{author}{\bibfnamefont{J.-S.~M.} \bibnamefont{Lehn}},
  \bibinfo{author}{\bibfnamefont{P.~V.} \bibnamefont{Rao}}, \bibnamefont{and}
  \bibinfo{author}{\bibfnamefont{R.~G.} \bibnamefont{Gordon}},
  \bibinfo{journal}{Appl. Phys. Lett.} \textbf{\bibinfo{volume}{89}},
  \bibinfo{eid}{133512} (\bibinfo{year}{2006}).

\bibitem[{\citenamefont{Thomas et~al.}(2007)\citenamefont{Thomas, Ehrhart,
  Roeckerath, van Elshocht, Rije, Luysberg, Boese, Schubert, Caymax, and
  Waser}}]{thomas}
\bibinfo{author}{\bibfnamefont{R.}~\bibnamefont{Thomas}},
  \bibinfo{author}{\bibfnamefont{P.}~\bibnamefont{Ehrhart}},
  \bibinfo{author}{\bibfnamefont{M.}~\bibnamefont{Roeckerath}},
  \bibinfo{author}{\bibfnamefont{S.}~\bibnamefont{van Elshocht}},
  \bibinfo{author}{\bibfnamefont{E.}~\bibnamefont{Rije}},
  \bibinfo{author}{\bibfnamefont{M.}~\bibnamefont{Luysberg}},
  \bibinfo{author}{\bibfnamefont{M.}~\bibnamefont{Boese}},
  \bibinfo{author}{\bibfnamefont{J.}~\bibnamefont{Schubert}},
  \bibinfo{author}{\bibfnamefont{M.}~\bibnamefont{Caymax}}, \bibnamefont{and}
  \bibinfo{author}{\bibfnamefont{R.}~\bibnamefont{Waser}}, \bibinfo{journal}{J.
  Electrochem. Soc.} \textbf{\bibinfo{volume}{154}}, \bibinfo{pages}{147 }
  (\bibinfo{year}{2007}).

\bibitem[{\citenamefont{Heeg et~al.}(2007)\citenamefont{Heeg, Roeckerath,
  Schubert, Zander, Buchal, Chen, Jia, Jia, Adamo, and Schlom}}]{heeg-2}
\bibinfo{author}{\bibfnamefont{T.}~\bibnamefont{Heeg}},
  \bibinfo{author}{\bibfnamefont{M.}~\bibnamefont{Roeckerath}},
  \bibinfo{author}{\bibfnamefont{J.}~\bibnamefont{Schubert}},
  \bibinfo{author}{\bibfnamefont{W.}~\bibnamefont{Zander}},
  \bibinfo{author}{\bibfnamefont{C.}~\bibnamefont{Buchal}},
  \bibinfo{author}{\bibfnamefont{H.~Y.} \bibnamefont{Chen}},
  \bibinfo{author}{\bibfnamefont{C.~L.} \bibnamefont{Jia}},
  \bibinfo{author}{\bibfnamefont{Y.}~\bibnamefont{Jia}},
  \bibinfo{author}{\bibfnamefont{C.}~\bibnamefont{Adamo}}, \bibnamefont{and}
  \bibinfo{author}{\bibfnamefont{D.~G.} \bibnamefont{Schlom}},
  \bibinfo{journal}{Appl. Phys. Lett.} \textbf{\bibinfo{volume}{90}},
  \bibinfo{eid}{192901} (\bibinfo{year}{2007}).

\bibitem[{\citenamefont{Wersing}(1996)}]{wersing}
\bibinfo{author}{\bibfnamefont{W.}~\bibnamefont{Wersing}},
  \bibinfo{journal}{Curr. Opin. Solid State Mater. Sci.}
  \textbf{\bibinfo{volume}{1}}, \bibinfo{pages}{715 } (\bibinfo{year}{1996}).

\bibitem[{\citenamefont{Vanderah}(2002)}]{terrell}
\bibinfo{author}{\bibfnamefont{T.~A.} \bibnamefont{Vanderah}},
  \bibinfo{journal}{Science} \textbf{\bibinfo{volume}{298}},
  \bibinfo{pages}{1182} (\bibinfo{year}{2002}).

\bibitem[{\citenamefont{Delugas et~al.}(2007)\citenamefont{Delugas, Fiorentini,
  Filippetti, and Pourtois}}]{delugas-th}
\bibinfo{author}{\bibfnamefont{P.}~\bibnamefont{Delugas}},
  \bibinfo{author}{\bibfnamefont{V.}~\bibnamefont{Fiorentini}},
  \bibinfo{author}{\bibfnamefont{A.}~\bibnamefont{Filippetti}},
  \bibnamefont{and} \bibinfo{author}{\bibfnamefont{G.}~\bibnamefont{Pourtois}},
  \bibinfo{journal}{Phys. Rev. B} \textbf{\bibinfo{volume}{75}},
  \bibinfo{eid}{115126} (\bibinfo{year}{2007}).

\bibitem[{\citenamefont{Vali}(2008)}]{vali-zr}
\bibinfo{author}{\bibfnamefont{R.}~\bibnamefont{Vali}}, \bibinfo{journal}{Solid
  State Commun.} \textbf{\bibinfo{volume}{145}}, \bibinfo{pages}{497 }
  (\bibinfo{year}{2008}).

\bibitem[{\citenamefont{Vali}(2009)}]{vali-hf}
\bibinfo{author}{\bibfnamefont{R.}~\bibnamefont{Vali}}, \bibinfo{journal}{Solid
  State Commun.} \textbf{\bibinfo{volume}{149}}, \bibinfo{pages}{519 }
  (\bibinfo{year}{2009}).

\bibitem[{EPA()}]{EPAPS_arxiv}
\bibinfo{note}{See
  \href{http://www.physics.rutgers.edu/~sinisa/highk/supp.pdf}{
  www.physics.rutgers.edu/\~{}sinisa/highk/supp.pdf} or
  \href{http://link.aps.org/supplemental/10.1103/PhysRevB.82.064101}{
  http://link.aps.org/supplemental/10.1103/PhysRevB.82.064101} for zone-center
  phonon frequencies, as well as the infrared activities for those modes that
  are infrared-active.}

\bibitem[{\citenamefont{Badie}(1978)}]{badie-1}
\bibinfo{author}{\bibfnamefont{J.}~\bibnamefont{Badie}}, \bibinfo{journal}{Rev.
  Int. Hautes Temp. R\' efact., Fr} \textbf{\bibinfo{volume}{15}},
  \bibinfo{pages}{183} (\bibinfo{year}{1978}).

\bibitem[{\citenamefont{Badie and Foex}(1978)}]{badie-2}
\bibinfo{author}{\bibfnamefont{J.}~\bibnamefont{Badie}} \bibnamefont{and}
  \bibinfo{author}{\bibfnamefont{M.}~\bibnamefont{Foex}}, \bibinfo{journal}{J.
  Solid State Chem.} \textbf{\bibinfo{volume}{26}}, \bibinfo{pages}{311}
  (\bibinfo{year}{1978}).

\bibitem[{\citenamefont{Coutures et~al.}(1980)\citenamefont{Coutures, Badie,
  Berjoan, Coutures, Flamand, and Rouanet}}]{coutures}
\bibinfo{author}{\bibfnamefont{J.-P.} \bibnamefont{Coutures}},
  \bibinfo{author}{\bibfnamefont{J.}~\bibnamefont{Badie}},
  \bibinfo{author}{\bibfnamefont{R.}~\bibnamefont{Berjoan}},
  \bibinfo{author}{\bibfnamefont{J.}~\bibnamefont{Coutures}},
  \bibinfo{author}{\bibfnamefont{R.}~\bibnamefont{Flamand}}, \bibnamefont{and}
  \bibinfo{author}{\bibfnamefont{A.}~\bibnamefont{Rouanet}},
  \bibinfo{journal}{High Temp. Sci.} \textbf{\bibinfo{volume}{13}},
  \bibinfo{pages}{331} (\bibinfo{year}{1980}).

\bibitem[{sch()}]{schubert}
\bibinfo{note}{J. Schubert (private communication).}

\bibitem[{\citenamefont{Berndt et~al.}(1975)\citenamefont{Berndt, Maier, and
  Keller}}]{berndt}
\bibinfo{author}{\bibfnamefont{U.}~\bibnamefont{Berndt}},
  \bibinfo{author}{\bibfnamefont{D.}~\bibnamefont{Maier}}, \bibnamefont{and}
  \bibinfo{author}{\bibfnamefont{C.}~\bibnamefont{Keller}},
  \bibinfo{journal}{J. Solid State Chem.} \textbf{\bibinfo{volume}{13}},
  \bibinfo{pages}{131} (\bibinfo{year}{1975}).

\bibitem[{\citenamefont{Rabenau}(1956)}]{labbo}
\bibinfo{author}{\bibfnamefont{A.}~\bibnamefont{Rabenau}}, \bibinfo{journal}{Z.
  Anorg. Allg. Chem.} \textbf{\bibinfo{volume}{288}}, \bibinfo{pages}{221}
  (\bibinfo{year}{1956}).

\bibitem[{\citenamefont{Ito et~al.}(2001)\citenamefont{Ito, Tezuka, and
  Hinatsu}}]{ito}
\bibinfo{author}{\bibfnamefont{K.}~\bibnamefont{Ito}},
  \bibinfo{author}{\bibfnamefont{K.}~\bibnamefont{Tezuka}}, \bibnamefont{and}
  \bibinfo{author}{\bibfnamefont{Y.}~\bibnamefont{Hinatsu}},
  \bibinfo{journal}{J. Solid State Chem.} \textbf{\bibinfo{volume}{157}},
  \bibinfo{pages}{173} (\bibinfo{year}{2001}).

\bibitem[{\citenamefont{Bharathy et~al.}(2009)\citenamefont{Bharathy, Fox,
  Mugavero, and zur Loye}}]{bharathy}
\bibinfo{author}{\bibfnamefont{M.}~\bibnamefont{Bharathy}},
  \bibinfo{author}{\bibfnamefont{A.}~\bibnamefont{Fox}},
  \bibinfo{author}{\bibfnamefont{S.}~\bibnamefont{Mugavero}}, \bibnamefont{and}
  \bibinfo{author}{\bibfnamefont{H.-C.} \bibnamefont{zur Loye}},
  \bibinfo{journal}{Solid State Sci.} \textbf{\bibinfo{volume}{11}},
  \bibinfo{pages}{651} (\bibinfo{year}{2009}).

\bibitem[{hee()}]{heeg-apl}
\bibinfo{note}{T. Heeg, K. Wiedenmann, M. Roeckerath, S. Coh, D. Vanderbilt, J.
  Schubert and D.~G. Schlom, unpublished.}

\bibitem[{\citenamefont{Goldschmidt}(1926)}]{goldschmidt}
\bibinfo{author}{\bibfnamefont{V.}~\bibnamefont{Goldschmidt}},
  \bibinfo{journal}{Naturwissenschaften} \textbf{\bibinfo{volume}{21}},
  \bibinfo{pages}{477} (\bibinfo{year}{1926}).

\bibitem[{\citenamefont{Woodward}(1997)}]{woodward-oct}
\bibinfo{author}{\bibfnamefont{P.~M.} \bibnamefont{Woodward}},
  \bibinfo{journal}{Acta Crystallogr. Sect. B} \textbf{\bibinfo{volume}{53}},
  \bibinfo{pages}{32} (\bibinfo{year}{1997}).

\bibitem[{\citenamefont{Glazer}(1972)}]{glazer-oct}
\bibinfo{author}{\bibfnamefont{A.~M.} \bibnamefont{Glazer}},
  \bibinfo{journal}{Acta Crystallogr. Sect. B} \textbf{\bibinfo{volume}{28}},
  \bibinfo{pages}{3384} (\bibinfo{year}{1972}).

\bibitem[{\citenamefont{Giannozzi et~al.}(2009)\citenamefont{Giannozzi, Baroni,
  Bonini, Calandra, Car, Cavazzoni, Ceresoli, Chiarotti, Cococcioni, Dabo
  et~al.}}]{QE-2009}
\bibinfo{author}{\bibfnamefont{P.}~\bibnamefont{Giannozzi}},
  \bibinfo{author}{\bibfnamefont{S.}~\bibnamefont{Baroni}},
  \bibinfo{author}{\bibfnamefont{N.}~\bibnamefont{Bonini}},
  \bibinfo{author}{\bibfnamefont{M.}~\bibnamefont{Calandra}},
  \bibinfo{author}{\bibfnamefont{R.}~\bibnamefont{Car}},
  \bibinfo{author}{\bibfnamefont{C.}~\bibnamefont{Cavazzoni}},
  \bibinfo{author}{\bibfnamefont{D.}~\bibnamefont{Ceresoli}},
  \bibinfo{author}{\bibfnamefont{G.~L.} \bibnamefont{Chiarotti}},
  \bibinfo{author}{\bibfnamefont{M.}~\bibnamefont{Cococcioni}},
  \bibinfo{author}{\bibfnamefont{I.}~\bibnamefont{Dabo}}, \bibnamefont{et~al.},
  \bibinfo{journal}{Journal of Physics: Condensed Matter}
  \textbf{\bibinfo{volume}{21}}, \bibinfo{pages}{395502 (19pp)}
  (\bibinfo{year}{2009}), \urlprefix\url{http://www.quantum-espresso.org}.

\bibitem[{\citenamefont{Perdew et~al.}(1996)\citenamefont{Perdew, Burke, and
  Ernzerhof}}]{perdew-pbe}
\bibinfo{author}{\bibfnamefont{J.~P.} \bibnamefont{Perdew}},
  \bibinfo{author}{\bibfnamefont{K.}~\bibnamefont{Burke}}, \bibnamefont{and}
  \bibinfo{author}{\bibfnamefont{M.}~\bibnamefont{Ernzerhof}},
  \bibinfo{journal}{Phys. Rev. Lett.} \textbf{\bibinfo{volume}{77}},
  \bibinfo{pages}{3865} (\bibinfo{year}{1996}).

\bibitem[{\citenamefont{Vanderbilt}(1985)}]{vanderbilt-uspp}
\bibinfo{author}{\bibfnamefont{D.}~\bibnamefont{Vanderbilt}},
  \bibinfo{journal}{Phys. Rev. B} \textbf{\bibinfo{volume}{32}},
  \bibinfo{pages}{8412} (\bibinfo{year}{1985}).

\bibitem[{\citenamefont{Monkhorst and Pack}(1976)}]{monkhorst-grid}
\bibinfo{author}{\bibfnamefont{H.~J.} \bibnamefont{Monkhorst}}
  \bibnamefont{and} \bibinfo{author}{\bibfnamefont{J.~D.} \bibnamefont{Pack}},
  \bibinfo{journal}{Phys. Rev. B} \textbf{\bibinfo{volume}{13}},
  \bibinfo{pages}{5188} (\bibinfo{year}{1976}).

\bibitem[{\citenamefont{Staroverov et~al.}(2004)\citenamefont{Staroverov,
  Scuseria, Tao, and Perdew}}]{gga-volume}
\bibinfo{author}{\bibfnamefont{V.~N.} \bibnamefont{Staroverov}},
  \bibinfo{author}{\bibfnamefont{G.~E.} \bibnamefont{Scuseria}},
  \bibinfo{author}{\bibfnamefont{J.}~\bibnamefont{Tao}}, \bibnamefont{and}
  \bibinfo{author}{\bibfnamefont{J.~P.} \bibnamefont{Perdew}},
  \bibinfo{journal}{Phys. Rev. B} \textbf{\bibinfo{volume}{69}},
  \bibinfo{pages}{075102} (\bibinfo{year}{2004}).

\bibitem[{\citenamefont{Pies and Weiss}(2006)}]{landolt-nitrides}
\bibinfo{author}{\bibfnamefont{W.}~\bibnamefont{Pies}} \bibnamefont{and}
  \bibinfo{author}{\bibfnamefont{A.}~\bibnamefont{Weiss}},
  \emph{\bibinfo{title}{Landolt-B\"ornstein - Group III Condensed Matter}}
  (\bibinfo{publisher}{Springer-Verlag}, \bibinfo{year}{2006}), vol.
  \bibinfo{volume}{7c1}, chap. \bibinfo{chapter}{VI.1.5.1}, pp.
  \bibinfo{pages}{14--34}.

\bibitem[{gdn()}]{gdn}
\bibinfo{note}{In the case of GdN the anomaly is more likely the result of the
  large spin splitting and strong ferromagnetism associated with the huge
  magnetic moment of the $4f^7$ configuration.}

\bibitem[{\citenamefont{Baroni et~al.}(2001)\citenamefont{Baroni, de~Gironcoli,
  Dal~Corso, and Giannozzi}}]{baroni-rmp}
\bibinfo{author}{\bibfnamefont{S.}~\bibnamefont{Baroni}},
  \bibinfo{author}{\bibfnamefont{S.}~\bibnamefont{de~Gironcoli}},
  \bibinfo{author}{\bibfnamefont{A.}~\bibnamefont{Dal~Corso}},
  \bibnamefont{and}
  \bibinfo{author}{\bibfnamefont{P.}~\bibnamefont{Giannozzi}},
  \bibinfo{journal}{Rev. Mod. Phys.} \textbf{\bibinfo{volume}{73}},
  \bibinfo{pages}{515} (\bibinfo{year}{2001}).

\bibitem[{\citenamefont{Cockayne and Burton}(2000)}]{cockayne}
\bibinfo{author}{\bibfnamefont{E.}~\bibnamefont{Cockayne}} \bibnamefont{and}
  \bibinfo{author}{\bibfnamefont{B.~P.} \bibnamefont{Burton}},
  \bibinfo{journal}{Phys. Rev. B} \textbf{\bibinfo{volume}{62}},
  \bibinfo{pages}{3735} (\bibinfo{year}{2000}).

\bibitem[{\citenamefont{Uecker et~al.}(2006)\citenamefont{Uecker, Wilke,
  Schlom, Velickov, Reiche, Polity, Bernhagen, and Rossberg}}]{uecker-1}
\bibinfo{author}{\bibfnamefont{R.}~\bibnamefont{Uecker}},
  \bibinfo{author}{\bibfnamefont{H.}~\bibnamefont{Wilke}},
  \bibinfo{author}{\bibfnamefont{D.}~\bibnamefont{Schlom}},
  \bibinfo{author}{\bibfnamefont{B.}~\bibnamefont{Velickov}},
  \bibinfo{author}{\bibfnamefont{P.}~\bibnamefont{Reiche}},
  \bibinfo{author}{\bibfnamefont{A.}~\bibnamefont{Polity}},
  \bibinfo{author}{\bibfnamefont{M.}~\bibnamefont{Bernhagen}},
  \bibnamefont{and} \bibinfo{author}{\bibfnamefont{M.}~\bibnamefont{Rossberg}},
  \bibinfo{journal}{J. Cryst. Growth} \textbf{\bibinfo{volume}{295}},
  \bibinfo{pages}{84} (\bibinfo{year}{2006}).

\bibitem[{\citenamefont{Uecker et~al.}(2008)\citenamefont{Uecker, Velickov,
  Klimm, Bertram, Bernhagen, Rabe, Albrecht, Fornari, and Schlom}}]{uecker-2}
\bibinfo{author}{\bibfnamefont{R.}~\bibnamefont{Uecker}},
  \bibinfo{author}{\bibfnamefont{B.}~\bibnamefont{Velickov}},
  \bibinfo{author}{\bibfnamefont{D.}~\bibnamefont{Klimm}},
  \bibinfo{author}{\bibfnamefont{R.}~\bibnamefont{Bertram}},
  \bibinfo{author}{\bibfnamefont{M.}~\bibnamefont{Bernhagen}},
  \bibinfo{author}{\bibfnamefont{M.}~\bibnamefont{Rabe}},
  \bibinfo{author}{\bibfnamefont{M.}~\bibnamefont{Albrecht}},
  \bibinfo{author}{\bibfnamefont{R.}~\bibnamefont{Fornari}}, \bibnamefont{and}
  \bibinfo{author}{\bibfnamefont{D.}~\bibnamefont{Schlom}},
  \bibinfo{journal}{J. Cryst. Growth} \textbf{\bibinfo{volume}{310}},
  \bibinfo{pages}{2649} (\bibinfo{year}{2008}).

\bibitem[{srz()}]{srzro}
\bibinfo{note}{The SrZrO$_3$, SrHfO$_3$, and LaScO$_3$ single crystals were
  grown by float-zone by Dima Souptel and Anatoly Balbashov of the Moscow Power
  Engineering Institute, Moscow, Russia. For details on the growth parameters
  see D. Souptel, G. Behr, and A.M. Balbashov, J. Cryst. Growth {\bf 236}, 583
  (2002).}

\bibitem[{\citenamefont{Newnham}(2005)}]{newnham}
\bibinfo{author}{\bibfnamefont{R.}~\bibnamefont{Newnham}},
  \emph{\bibinfo{title}{Properties of Materials: Anisotropy, Symmetry,
  Structure}} (\bibinfo{publisher}{Oxford University Press, Oxford},
  \bibinfo{year}{2005}).

\bibitem[{\citenamefont{Gesing et~al.}(2009)\citenamefont{Gesing, Uecker, and
  Buhl}}]{gesing}
\bibinfo{author}{\bibfnamefont{T.~M.} \bibnamefont{Gesing}},
  \bibinfo{author}{\bibfnamefont{R.}~\bibnamefont{Uecker}}, \bibnamefont{and}
  \bibinfo{author}{\bibfnamefont{J.-C.} \bibnamefont{Buhl}},
  \bibinfo{journal}{Z. Kristallogr. NCS} \textbf{\bibinfo{volume}{224}},
  \bibinfo{pages}{365} (\bibinfo{year}{2009}).

\bibitem[{\citenamefont{Anderson et~al.}(1993)\citenamefont{Anderson,
  Greenwood, Taylor, and Poeppelmeier}}]{anderson-bord}
\bibinfo{author}{\bibfnamefont{M.~T.} \bibnamefont{Anderson}},
  \bibinfo{author}{\bibfnamefont{K.~B.} \bibnamefont{Greenwood}},
  \bibinfo{author}{\bibfnamefont{G.~A.} \bibnamefont{Taylor}},
  \bibnamefont{and} \bibinfo{author}{\bibfnamefont{K.~R.}
  \bibnamefont{Poeppelmeier}}, \bibinfo{journal}{Prog. Sol. St. Chem.}
  \textbf{\bibinfo{volume}{22}}, \bibinfo{pages}{197 } (\bibinfo{year}{1993}).

\bibitem[{\citenamefont{Liferovich and Mitchell}(2004)}]{liferovich-exp}
\bibinfo{author}{\bibfnamefont{R.~P.} \bibnamefont{Liferovich}}
  \bibnamefont{and} \bibinfo{author}{\bibfnamefont{R.~H.}
  \bibnamefont{Mitchell}}, \bibinfo{journal}{J. Phys. Chem. Solids}
  \textbf{\bibinfo{volume}{177}}, \bibinfo{pages}{2188 }
  (\bibinfo{year}{2004}).

\bibitem[{\citenamefont{Velickov et~al.}(2007)\citenamefont{Velickov,
  Kahlenberg, Bertram, and Bernhagen}}]{velickov-exp}
\bibinfo{author}{\bibfnamefont{B.}~\bibnamefont{Velickov}},
  \bibinfo{author}{\bibfnamefont{V.}~\bibnamefont{Kahlenberg}},
  \bibinfo{author}{\bibfnamefont{R.}~\bibnamefont{Bertram}}, \bibnamefont{and}
  \bibinfo{author}{\bibfnamefont{M.}~\bibnamefont{Bernhagen}},
  \bibinfo{journal}{Z. Kristallogr.} \textbf{\bibinfo{volume}{9}},
  \bibinfo{pages}{466} (\bibinfo{year}{2007}).

\bibitem[{\citenamefont{Kennedy
  et~al.}(1999{\natexlab{a}})\citenamefont{Kennedy, Howard, and
  Chakoumakos}}]{kennedy-exp}
\bibinfo{author}{\bibfnamefont{B.~J.} \bibnamefont{Kennedy}},
  \bibinfo{author}{\bibfnamefont{C.~J.} \bibnamefont{Howard}},
  \bibnamefont{and} \bibinfo{author}{\bibfnamefont{B.~C.}
  \bibnamefont{Chakoumakos}}, \bibinfo{journal}{Phys. Rev. B}
  \textbf{\bibinfo{volume}{59}}, \bibinfo{pages}{4023}
  (\bibinfo{year}{1999}{\natexlab{a}}).

\bibitem[{\citenamefont{Kennedy
  et~al.}(1999{\natexlab{b}})\citenamefont{Kennedy, Howard, and
  Chakoumakos}}]{kennedy2-exp}
\bibinfo{author}{\bibfnamefont{B.~J.} \bibnamefont{Kennedy}},
  \bibinfo{author}{\bibfnamefont{C.~J.} \bibnamefont{Howard}},
  \bibnamefont{and} \bibinfo{author}{\bibfnamefont{B.~C.}
  \bibnamefont{Chakoumakos}}, \bibinfo{journal}{Phys. Rev. B}
  \textbf{\bibinfo{volume}{60}}, \bibinfo{pages}{2972}
  (\bibinfo{year}{1999}{\natexlab{b}}).

\bibitem[{\citenamefont{Velickov et~al.}(2008)\citenamefont{Velickov,
  Kahlenberg, Bertram, and Uecker}}]{velickov-tbsco}
\bibinfo{author}{\bibfnamefont{B.}~\bibnamefont{Velickov}},
  \bibinfo{author}{\bibfnamefont{V.}~\bibnamefont{Kahlenberg}},
  \bibinfo{author}{\bibfnamefont{R.}~\bibnamefont{Bertram}}, \bibnamefont{and}
  \bibinfo{author}{\bibfnamefont{R.}~\bibnamefont{Uecker}},
  \bibinfo{journal}{Acta Crystallogr. Sect. E} \textbf{\bibinfo{volume}{64}},
  \bibinfo{pages}{i79} (\bibinfo{year}{2008}).

\bibitem[{\citenamefont{Ruiz-Trejo et~al.}(1999)\citenamefont{Ruiz-Trejo,
  Islam, and Kilner}}]{ruiz}
\bibinfo{author}{\bibfnamefont{E.}~\bibnamefont{Ruiz-Trejo}},
  \bibinfo{author}{\bibfnamefont{M.~S.} \bibnamefont{Islam}}, \bibnamefont{and}
  \bibinfo{author}{\bibfnamefont{J.~A.} \bibnamefont{Kilner}},
  \bibinfo{journal}{Solid State Ionics} \textbf{\bibinfo{volume}{123}},
  \bibinfo{pages}{121 } (\bibinfo{year}{1999}).

\bibitem[{\citenamefont{Levin et~al.}(2003)\citenamefont{Levin, Amos, Bell,
  Farber, Vanderah, Roth, and Toby}}]{cazro}
\bibinfo{author}{\bibfnamefont{I.}~\bibnamefont{Levin}},
  \bibinfo{author}{\bibfnamefont{T.~G.} \bibnamefont{Amos}},
  \bibinfo{author}{\bibfnamefont{S.~M.} \bibnamefont{Bell}},
  \bibinfo{author}{\bibfnamefont{L.}~\bibnamefont{Farber}},
  \bibinfo{author}{\bibfnamefont{T.~A.} \bibnamefont{Vanderah}},
  \bibinfo{author}{\bibfnamefont{R.~S.} \bibnamefont{Roth}}, \bibnamefont{and}
  \bibinfo{author}{\bibfnamefont{B.~H.} \bibnamefont{Toby}},
  \bibinfo{journal}{J. Solid State Chem.} \textbf{\bibinfo{volume}{175}},
  \bibinfo{pages}{170 } (\bibinfo{year}{2003}), ISSN \bibinfo{issn}{0022-4596}.

\bibitem[{ben()}]{bennett}
\bibinfo{note}{J.~W. Bennett, private communication.}

\bibitem[{\citenamefont{Bennett et~al.}(2008)\citenamefont{Bennett, Grinberg,
  and Rappe}}]{bennett-cazro}
\bibinfo{author}{\bibfnamefont{J.~W.} \bibnamefont{Bennett}},
  \bibinfo{author}{\bibfnamefont{I.}~\bibnamefont{Grinberg}}, \bibnamefont{and}
  \bibinfo{author}{\bibfnamefont{A.~M.} \bibnamefont{Rappe}},
  \bibinfo{journal}{Chem. Mater} \textbf{\bibinfo{volume}{20}},
  \bibinfo{pages}{5134} (\bibinfo{year}{2008}).

\bibitem[{\citenamefont{Arnold et~al.}(2009)\citenamefont{Arnold, Knight,
  Morrison, and Lightfoot}}]{arnold}
\bibinfo{author}{\bibfnamefont{D.~C.} \bibnamefont{Arnold}},
  \bibinfo{author}{\bibfnamefont{K.~S.} \bibnamefont{Knight}},
  \bibinfo{author}{\bibfnamefont{F.~D.} \bibnamefont{Morrison}},
  \bibnamefont{and}
  \bibinfo{author}{\bibfnamefont{P.}~\bibnamefont{Lightfoot}},
  \bibinfo{journal}{Phys. Rev. Lett.} \textbf{\bibinfo{volume}{102}},
  \bibinfo{pages}{027602} (\bibinfo{year}{2009}).

\bibitem[{\citenamefont{Haumont et~al.}(2008)\citenamefont{Haumont, Kornev,
  Lisenkov, Bellaiche, Kreisel, and Dkhil}}]{haumont}
\bibinfo{author}{\bibfnamefont{R.}~\bibnamefont{Haumont}},
  \bibinfo{author}{\bibfnamefont{I.~A.} \bibnamefont{Kornev}},
  \bibinfo{author}{\bibfnamefont{S.}~\bibnamefont{Lisenkov}},
  \bibinfo{author}{\bibfnamefont{L.}~\bibnamefont{Bellaiche}},
  \bibinfo{author}{\bibfnamefont{J.}~\bibnamefont{Kreisel}}, \bibnamefont{and}
  \bibinfo{author}{\bibfnamefont{B.}~\bibnamefont{Dkhil}},
  \bibinfo{journal}{Phys. Rev. B} \textbf{\bibinfo{volume}{78}},
  \bibinfo{pages}{134108} (\bibinfo{year}{2008}).

\bibitem[{\citenamefont{Berkstresser et~al.}(1993)\citenamefont{Berkstresser,
  Valentino, and Brandle}}]{berkstresser}
\bibinfo{author}{\bibfnamefont{G.~W.} \bibnamefont{Berkstresser}},
  \bibinfo{author}{\bibfnamefont{A.~J.} \bibnamefont{Valentino}},
  \bibnamefont{and} \bibinfo{author}{\bibfnamefont{C.~D.}
  \bibnamefont{Brandle}}, \bibinfo{journal}{J. Cryst. Growth}
  \textbf{\bibinfo{volume}{128}}, \bibinfo{pages}{684} (\bibinfo{year}{1993}).

\bibitem[{\citenamefont{Ovanesyan et~al.}(1999)\citenamefont{Ovanesyan,
  Petrosyan, Shirinyan, Pedrini, and Zhang}}]{ovanesyan}
\bibinfo{author}{\bibfnamefont{K.~L.} \bibnamefont{Ovanesyan}},
  \bibinfo{author}{\bibfnamefont{A.}~\bibnamefont{Petrosyan}},
  \bibinfo{author}{\bibfnamefont{G.~O.} \bibnamefont{Shirinyan}},
  \bibinfo{author}{\bibfnamefont{C.}~\bibnamefont{Pedrini}}, \bibnamefont{and}
  \bibinfo{author}{\bibfnamefont{L.}~\bibnamefont{Zhang}}, \bibinfo{journal}{J.
  Cryst. Growth} \textbf{\bibinfo{volume}{198}}, \bibinfo{pages}{497}
  (\bibinfo{year}{1999}).

\bibitem[{spo()}]{spoint}
\bibinfo{note}{The eigendisplacement of this unstable mode is very close to
  that of the $S$-point $(\frac{1}{2} \frac{1}{2} 0)$ phonon of the 20-atom
  \pbnm\ structure having a frequency of 39~cm$^{-1}$. However, this phonon
  remains inactive in the \pbnm\ structure because it does not appear at the
  $\Gamma$ point.}

\bibitem[{mul()}]{muller}
\bibinfo{note}{D. A. Muller, private communication.}

\end{thebibliography}

\end{document}